\newif\ifconfver
\confverfalse      

\newif\ifcutshort      
\cutshorttrue

\newif\ifcutshortlvltwo  
\cutshortlvltwofalse

\ifconfver
\documentclass[10pt,twocolumn,twoside]{IEEEtran}
\else
\documentclass[11pt,draftcls,onecolumn]{IEEEtran}
\fi

\usepackage{array}
\usepackage{cite}
\usepackage{calc}
\usepackage{caption}
\usepackage{graphicx}
\usepackage{psfrag}
\usepackage[tight,footnotesize]{subfigure}
\usepackage{stfloats}
\usepackage{url}
\usepackage{subfig}
\usepackage{longtable}
\usepackage{amsmath,amsfonts,amssymb,amsbsy,bm,paralist,theorem,ifthen,color}
\usepackage{algorithmic,algorithm}
\usepackage[dvips]{epsfig}
\usepackage[dvips]{graphicx}

\newcommand\Vc{\ensuremath{{\mathcal{V}}}}

\newcommand\Lc{\ensuremath{{\mathcal{L}}}}

\newcommand\Ac{\ensuremath{{\mathcal{A}}}}
\newcommand\Nc{\ensuremath{{\mathcal{N}}}}
\newcommand\xb{\ensuremath{{\bm x}}}

\newcommand\wb{\ensuremath{{\bm w}}}

\newcommand\ssb{\ensuremath{{\bm s}}}

\newcommand\Ab{\ensuremath{{\bm A}}}
\newcommand\ab{\ensuremath{{\bm a}}}
\newcommand\Bb{\ensuremath{{\bm B}}}
\newcommand\bb{\ensuremath{{\bm b}}}

\newcommand\lambdab{\ensuremath{{\bm \lambda}}}
\newcommand\nub{\ensuremath{{\bm \nu}}}

\newcommand\zerob{\ensuremath{{\bm 0}}}

\newtheorem{Lemma}{Lemma}

\newtheorem{Theorem}{Theorem}

\newtheorem{Corollary}{Corollary}

\newtheorem{Assumption}{Assumption}

\ifconfver

\else

\fi

\begin{document}
	
	\bibliographystyle{IEEEtran}
	
	\title{Asynchronous ADMM for Distributed Optimization with Heterogeneous Processors}
	\title{Asynchronous Distributed ADMM for Large-Scale Optimization- Part I: \\Algorithm and Convergence Analysis}

	\ifconfver \else {\linespread{1.1} \rm \fi
		
		\author{\vspace{0.8cm}Tsung-Hui Chang$^\star$, Mingyi Hong$^\dag$, Wei-Cheng Liao$^\S$ and Xiangfeng Wang$^\ddag$\\
			\thanks{Part of this work was submitted to { IEEE ICASSP,  Shanghai, China, March 20-25, 2016 \cite{ChangNIPS15}.}
				Tsung-Hui Chang is supported by NSFC, China, Grant No. 61571385.  Mingyi Hong is supported by NFS Grant No. CCF-1526078 , and AFOSR, Grant No. 15RT0767. Xiangfeng Wang is supported by Shanghai YangFan No. 15YF1403400 and NSFC No. 11501210.}
			\thanks{$^\star$Tsung-Hui Chang is the corresponding author. Address:
				School of Science and Engineering, The Chinese University of Hong Kong, Shenzhen, China 518172. E-mail:
				tsunghui.chang@ieee.org. }
			\thanks{$^\dag$Mingyi Hong is with Department of Industrial and Manufacturing Systems Engineering, Iowa State University, Ames, 50011, USA, E-mail: mingyi@iastate.edu}
			\thanks{$^\dag$Wei-Cheng Liao is with Department of Electrical and Computer Engineering, University of Minnesota, Minneapolis, MN 55455, USA, E-mail: liaox146@umn.edu}
			\thanks{$^\ddag$Xiangfeng Wang is with Shanghai Key Lab for Trustworthy Computing, School of Computer Science and Software Engineering, East China Normal University, Shanghai, 200062, China, E-mail: xfwang@sei.ecnu.edu.cn}
		}
		\maketitle
		\vspace{-0.5cm}
		\begin{abstract}
			Aiming at solving large-scale optimization problems, this paper studies distributed { optimization} methods based on the alternating direction method of multipliers (ADMM).
			By formulating the { optimization} problem as a consensus problem, the ADMM can be used to solve the consensus problem in a fully parallel fashion over a computer network with a star topology. However, traditional synchronized computation does not scale well with the problem size, as the speed of the algorithm is limited by the slowest workers. This is particularly true in a heterogeneous network where the computing nodes experience different computation and communication delays.
			In this paper, we propose an asynchronous distributed ADMM {(AD-ADMM)} which can effectively improve the time efficiency of distributed optimization. Our main interest lies in analyzing the convergence conditions of the AD-ADMM, under the popular \emph{partially asynchronous} model, which is defined based on a maximum tolerable delay of the network. Specifically, by considering general and possibly non-convex cost functions, we show that the AD-ADMM is guaranteed to converge to the set of Karush-Kuhn-Tucker (KKT) points as long as the algorithm parameters are chosen appropriately according to the network delay. We further illustrate that the asynchrony of the ADMM has to be handled with care, as slightly modifying the implementation of the AD-ADMM can jeopardize the algorithm convergence, even under { the} standard convex setting.
		\end{abstract}
		
		\ifconfver \else
		\newpage
		\fi
		
		\ifconfver \else \IEEEpeerreviewmaketitle} \fi

	\vspace{-0.3cm}
	\section{Introduction}\label{sec: intro}
	\vspace{-0.0cm}
	\subsection{ Background}
	Scaling up optimization algorithms for future data-intensive applications calls for efficient distributed and parallel implementations, so that modern multi-core high performance computing technologies can be fully utilized \cite{Cevher14SPM,BK:Bekkerman12,BertsekasADMM}. 
	In this work, we are interested in developing distributed optimization methods for solving the following optimization problem
	\begin{align}\label{eqn: original problem}
		\min_{\substack{\xb\in \mathbb{R}^n}}~ &\sum_{i=1}^N f_i(\xb) + h(\xb),
	\end{align}
	where each $f_i :\mathbb{R}^{n} \rightarrow \mathbb{R}$ is a (smooth) cost function; $h:\mathbb{R}^{n} \rightarrow  \mathbb{R}\cup\{\infty\}$ is a convex ({ proper and} lower semi-continuous) but possibly non-smooth regularization function. The latter is used to impose desired structures on the solution (e.g., sparsity) and/or used to enforce certain constraints. 
	Problem \eqref{eqn: original problem} includes as special cases many important statistical learning problems such as the { LASSO problem \cite{Tibshirani96},} logistic regression (LR) problem \cite{Liu2009}, support vector machine (SVM) \cite{Hastie2001Book} and the sparse principal component analysis (PCA) problem \cite{Richtarikspca12}. In this paper, we focus on solving large-scale instances of these learning problems with either a large number of training samples or a large number of features ($n$ is large) \cite{BK:Bekkerman12}. These are typical data-intensive machine learning scenarios in which the data sets are often distributedly located in a few computing nodes. Traditional centralized optimization methods, therefore, fails to scale well due to their inability to handle distributed data sets and computing resources.  
	
\begin{figure}[t]\centering
		\resizebox{0.45\textwidth}{!}{
			\includegraphics{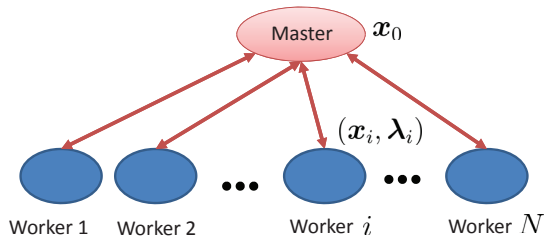}}
		\caption{A star computer cluster with one master and $N$ workers.}
		\vspace{-0.5cm}\label{fig: masterslave network}
	\end{figure}

	Our goal is to develop efficient distributed optimization algorithms over a computer network with a star topology, in which a master node coordinates the computation of a set of distributed workers (see Figure \ref{fig: masterslave network} for illustration).
	Such star topology represents a common architecture for distributed computing, therefore it has been used widely in distributed optimization \cite{BertsekasADMM,BoydADMM11,NiuNIPS11,AgarwalNIPS11,MLiPS14,MLiNIPS14,LiuWright15, meisam14nips, scutari13decomposition}. For example, under the star topology, references \cite{NiuNIPS11,AgarwalNIPS11} presented distributed stochastic gradient descent (SGD) methods, references \cite{MLiPS14,MLiNIPS14} parallelized the proximal gradient (PG) methods, while references \cite{LiuWright15,meisam14nips, scutari13decomposition,DaneshmandTSP15} parallelized the block coordinate descent (BCD) method. In these works, the distributed workers iteratively calculate the gradients related to their local data, while the master collects such information from the workers to perform SGD, PG or BCD updates.
	
	However, when scaling up these distributed algorithms, node synchronization becomes an important issue.
	Specifically, under the synchronous protocol, the master is triggered at each iteration only if it receives the required information from all the distributed workers.
	On the one hand, such synchronization is beneficial to make the algorithms well behaved; on the other hand, however, the speed of the algorithms would be limited by the ``slowest" worker especially when the workers have different computation and communication delays. To address such dilemma, a few recent works \cite{NiuNIPS11,AgarwalNIPS11,MLiPS14,MLiNIPS14,LiuWright15} have introduced ``asynchrony" into the distributed algorithms, which allows the master to perform updates when not all, but a small subset of workers have returned their gradient information. The asynchronous updates would cause ``delayed" gradient information.
	A few algorithmic tricks such as delay-dependent step-size selection have been introduced to ensure that the staled gradient information does not destroy the stability of the algorithm. In practice, such asynchrony does make a big difference.
	As has been consistently reported in \cite{NiuNIPS11,AgarwalNIPS11,MLiPS14,MLiNIPS14,LiuWright15}, under such an asynchronous protocol, the computation time can decrease almost linearly with the number of workers.

	\vspace{-0.2cm}
	\subsection{ Related Works}
	
	A different approach for distributed and parallel optimization is based on the alternating direction method of multipliers (ADMM) \cite[Section 7.1.1]{BoydADMM11}. In the distributed ADMM, the original learning problem is partitioned into $N$ subproblems, each containing a subset of training samples or the learning parameters. 
	At each iteration, the workers solve the subproblems and send the up-to-date variable information to the master, who summarizes { this information} and broadcasts the result to the workers. In this way, a given large-scale learning problem can be solved in a parallel and distributed fashion. Notably, other than the standard convex setting \cite{BoydADMM11}, the recent analysis in \cite{Hong15noncvxadmm} has shown that such distributed ADMM is provably convergent to a Karush-Kuhn-Tucker (KKT) point even for non-convex problems.

	{ Recently,  the synchronous distributed ADMM \cite{BoydADMM11,Hong15noncvxadmm} has been extended to the asynchronous setting, similar to \cite{NiuNIPS11,AgarwalNIPS11,MLiPS14,MLiNIPS14,LiuWright15}.
		Specifically, reference \cite{Zhang14ACADMM} has considered a version of AD-ADMM with bounded delay assumption and studied its theoretical and numerical performances.} However, only convex cases are considered in \cite{Zhang14ACADMM}. 
	Reference \cite{Hong15noncvxadmm2} has studied another version of AD-ADMM for non-convex problems, 
	which considers inexact subproblem updates and, similar to \cite{NiuNIPS11,AgarwalNIPS11,MLiPS14,MLiNIPS14,LiuWright15}, the workers compute gradient information only. { This type of distributed optimization schemes, however, may not fully utilize the computation powers of distributed nodes.
		Besides, due to inexact update, such schemes usually require more iterations to converge and thus may have higher communication overhead.}
	References{ \cite{IutzelerCDC13,ErminWei2013arxiv,MotaDADMM2013}} have respectively considered asynchronous ADMM methods for decentralized optimization over networks. These works consider network topologies beyond the star network, but their definition of asynchrony is different from what we propose here. Specifically, the asynchrony in \cite{IutzelerCDC13} lies in that, at each iteration, the nodes are randomly activated to perform variable update.
	The method presented in \cite{ErminWei2013arxiv} further allows that the communications between nodes can succeed or fail randomly. It is shown in \cite{ErminWei2013arxiv} that such asynchronous ADMM can converge in a probability-one sense, provided that the nodes and communication links satisfy certain statistical assumption. {Reference \cite{MotaDADMM2013}  has considered an {asynchronous} dual ADMM method. The asynchrony is in the sense that the nodes are partitioned into groups based on certain coloring scheme and only one group of nodes update variable in each iteration. }
	

	
	
	\vspace{-0.2cm}
	\subsection{ Contributions}

	{In this paper\footnote{In contrast to the conference paper \cite{ChangNIPS15},  the current paper presents detailed proofs of theorems and more simulation results.},} we generalize the state-of-the-art synchronous distributed ADMM \cite{BoydADMM11,Hong15noncvxadmm} to the asynchronous setting. Like \cite{NiuNIPS11,AgarwalNIPS11,MLiPS14,MLiNIPS14,LiuWright15,Zhang14ACADMM,Hong15noncvxadmm2},
	the asynchronous distributed ADMM (AD-ADMM) algorithm developed in this paper gives the master the freedom of
	making updates only based on variable information from a partial set of workers, which further improves the computation efficiency of the distributed ADMM.
	
	Theoretically, we show that, for general and possibly non-convex problems in the form of \eqref{eqn: original problem}, the AD-ADMM converges to the set of KKT points if the algorithm parameters are chosen appropriately according to the maximum network delay.
	{Our results differ significantly from the existing works \cite{Zhang14ACADMM,IutzelerCDC13,ErminWei2013arxiv} which are all developed for convex problems.  }
	Therefore, the analysis and algorithm proposed here are applicable not only to standard convex learning problems but also to important non-convex problems such as the sparse PCA problem \cite{Richtarikspca12} and matrix factorization problems \cite{LingmcICASSP12}. To the best of our knowledge, except the inexact version in \cite{Hong15noncvxadmm2}, this is the first time that the distributed ADMM is rigorously shown to be convergent for non-convex problems under the asynchronous protocol.
	Moreover, unlike \cite{Zhang14ACADMM,IutzelerCDC13,ErminWei2013arxiv} where the convergence analyses all rely
	on certain statistical assumption on the nodes/workers, our convergence analysis is deterministic and characterizes the worst-case convergence conditions of the AD-ADMM under a bounded delay assumption only.
	{Furthermore, we demonstrate that the asynchrony of ADMM has to be handled with care -- as a slight modification of the algorithm may lead to completely different convergence conditions and even destroy the convergence of ADMM for convex problems. Some numerical results are presented to support our theoretical claims.

		In the companion paper
		\cite{ChangAsyncadmm15_p2}, the linear convergence conditions of the AD-ADMM is further analyzed. In addition, the numerical performance of the AD-ADMM is examined by
		solving a large-scale LR problem on a high-performance computer cluster.}

	{\bf Synopsis:} Section \ref{sec: applications and D-ADMM} presents the applications of problem \eqref{eqn: original problem} and reviews the distributed ADMM in \cite{BoydADMM11}.
	The proposed AD-ADMM and its convergence conditions are presented in Section \ref{sec: async D-ADMM}. Comparison of the proposed AD-ADMM with an alternative scheme is presented in Section \ref{sec: alternative scheme}. Some simulation results are presented in Section \ref{sec: simu}. Finally, concluding remarks are given in Section \ref{sec: conclusions}.

	

	\section{Applications and Distributed ADMM}\label{sec: applications and D-ADMM}
	
	\subsection{Applications}\label{subsec: apps}
	We target at solving problem \eqref{eqn: original problem} over a star computer network (cluster) with one master node and $N$ workers/slaves, as illustrated in Figure \ref{fig: masterslave network}. 
	Such distributed optimization approach is extremely useful in modern big data applications \cite{BK:Bekkerman12}.
	For example, let us consider the following regularized empirical risk minimization problem \cite{Hastie2001Book}
	\begin{align}\label{eqn: risk min prob}
		\min_{\substack{\wb\in \mathbb{R}^n}}~ &\sum_{j=1}^m \ell(\ab_j^T\wb,y_j) + \Omega(\wb),
	\end{align}
	where $m$ is the number of training samples and $\ell(\ab_j^T\wb,y_j)$ is a loss function (e.g., regression or classification error) that depends on the training sample
	$\ab_j\in \mathbb{R}^n$, label $y_j$ and the parameter vector $\wb \in \mathbb{R}^n$. Here, $n$ denotes the dimension of the parameters (features); $\Omega(\wb)$ is an {appropriate convex regularizer}.
	Problem \eqref{eqn: risk min prob} is one of the most important problems in signal processing and statistical learning, which includes the LASSO problem \cite{Tibshirani05_fusedLASSO}, LR \cite{Liu2009}, SVM \cite{Hastie2001Book} and the sparse PCA problem \cite{Richtarikspca12}, to name a few.
	Obviously, solving \eqref{eqn: risk min prob} can be challenging when the number of training samples is very large. In that case, it is natural to split the training samples across the computer cluster and resort to a distributed optimization approach. Suppose that the $m$ training samples are uniformly distributed and stored by the $N$ workers, with each node $i$ getting $q_i = \llcorner m/N \lrcorner$ samples. By defining $f_i(\wb)\triangleq \sum_{j=(i-1)q_i+1}^{i q_i}
	\ell(\ab_j^T\wb,y_j)$, $i=1,\ldots,N$, and $h(\wb)\triangleq\Omega(\wb)$, it is clear that \eqref{eqn: risk min prob} is an instance of \eqref{eqn: original problem}.

	When the number of training samples is moderate but the dimension of the parameters is very large ($n \gg m$), problem \eqref{eqn: risk min prob} is also challenging to solve. By \cite[Section 7.3]{BoydADMM11}, one can instead consider the Lagrangian dual problem of \eqref{eqn: risk min prob} provided that \eqref{eqn: risk min prob}
	has {zero duality gap.} Specifically, let the training matrix $\Ab\triangleq [\ab_1,\ldots,\ab_m]^T \in \mathbb{R}^{m\times n}$ be partitioned as $\Ab=[\Ab_1,\ldots,\Ab_N]$, and let
	the parameter vector $\wb$ be partitioned conformally as $\wb=[\wb_1^T,\ldots,\wb_N^T]^T$; moreover, assume that $\Omega$ is separable as $\Omega(\wb)=\sum_{i=1}^N \Omega_i(\wb_i)$. Then, following \cite[Section 7.3]{BoydADMM11}, one can obtain the dual problem of \eqref{eqn: risk min prob} as
	\begin{align}\label{eqn: risk min prob dual}
		\min_{\substack{\nub\in \mathbb{R}^m}}~ & \sum_{i=1}^N \Omega_i^*(\Ab_i^T\nub) +\Phi^*(\nub),
	\end{align}
	where $\nub\triangleq[\nu_1,\ldots,\nu_m]^T$ is a dual variable, $\Phi^*(\nub)=\sum_{j=1}^m \ell^*(\nu_j,y_j)$, and $\ell^*$ and $\Omega_i^*$ are respectively the conjugate functions of $\ell$ and $\Omega_i$. Note that \eqref{eqn: risk min prob dual} is equivalent to splitting the $n$ parameters across the $N$ workers. Clearly, problem  \eqref{eqn: risk min prob dual} is an instance of \eqref{eqn: original problem}.
	
	It is interesting to mention that many emerging problems in smart power grid {can also be formulated as} problem \eqref{eqn: original problem}; see, for example, the power state estimation problem considered in \cite{ZhangPES15}
	is solved by employing the distributed ADMM. The energy management problems (i.e., demand response) in \cite{ChangTAC2014,JooTSG13,DallAneseTSG13} can potentially be handled by the distributed ADMM as well.
	

	\subsection{Distributed ADMM}\label{subsec: D-ADMM}
	
	In this section, we present the distributed ADMM \cite{BertsekasADMM,BoydADMM11} for solving problem \eqref{eqn: original problem}.
	Let us consider the following consensus formulation of problem \eqref{eqn: original problem}
	\begin{subequations}\label{eqn: consensus problem}
		\begin{align}
			\min_{\substack{\xb_0,\xb_i\in \mathbb{R}^n, \\ i=1,\ldots,N}} &\quad \sum_{i=1}^N f_i(\xb_i) + h(\xb_0)
			\\
			\text{s.t.}& \quad  \xb_i =\xb_0,\quad \forall i=1,\ldots,N.\label{eqn: consensus problem C1}
		\end{align}
	\end{subequations}
	In \eqref{eqn: consensus problem}, the $N+1$ variables $\xb_i$, $i=0,1,\ldots,N$, are subject to the consensus constraint in \eqref{eqn: consensus problem C1}, i.e., $\xb_0=\xb_1=\cdots=\xb_N$. Thus, problem \eqref{eqn: consensus problem} is equivalent to \eqref{eqn: original problem}.
	
	It has been shown that such a consensus problem can be efficiently solved by the ADMM \cite{BoydADMM11}. To describe this method,
	let $\lambdab\in \mathbb{R}^n$ denote the Lagrange dual variable associated with constraint \eqref{eqn: consensus problem C1} and define the following augmented Lagrangian function
	\begin{align}\label{eqn: Lc0}
		\Lc_{\rho}(\xb,\xb_0, \lambdab)&=
		\sum_{i=1}^N f_i(\xb_i) + h(\xb_0) \notag \\
		&+\sum_{i=1}^N \lambdab_i^T(\xb_i -\xb_0)+\frac{\rho}{2}\sum_{i=1}^N\|\xb_i -\xb_0\|^2,
	\end{align}
	where $\xb\triangleq[\xb_1^T,\ldots,\xb_N^T]^T$, $\lambdab\triangleq[\lambdab_1^T,\ldots,\lambdab_N^T]^T$ and
	$\rho>0$ is a penalty parameter. According to \cite{BertsekasADMM}, the standard synchronous ADMM iteratively updates the primal variables $\xb_i,i=0,1,\ldots,N,$ by minimizing \eqref{eqn: Lc0} in a (one-round) Gauss-Seidel fashion, followed by updating the dual variable $\lambdab$ using an approximate gradient ascent method. The ADMM algorithm for solving \eqref{eqn: consensus problem} is presented in Algorithm \ref{table: sync cadmm},
	
	\begin{algorithm}[h!]
		\caption{(Synchronous) Distributed ADMM for \eqref{eqn: consensus problem} \cite{BoydADMM11}}
		\begin{algorithmic}[1]\label{table: sync cadmm}
			\STATE {\bf Given} initial variables
			$\xb^{0}$ and $\lambdab^{0}$; set $\xb_0^0=\xb^{0}$ and $k=0.$
			\REPEAT
			\STATE  {\bf update}
			\begin{align}
				\!\!\!\!\xb^{k+1}_0\! &\!=\! \arg\min_{\xb_0 \in \mathbb{R}^n}\!  \textstyle \bigg\{h(\xb_0)-  \xb_0^T\sum_{i=1}^N \lambdab_i^{k} \notag \\
				&\textstyle~~~~~~~~~~~~~~~~~~+\frac{\rho}{2}\sum_{i=1}^N\|\xb_i^{k}-\xb_0\|^2\bigg\},\label{eqn: sync cadmm s2}
				\\
				\xb^{k+1}_i & \!=\!\arg\min_{\xb_i\in \mathbb{R}^n}~ \textstyle f_i(\xb_i)+\xb_i^T\lambdab_i^k +\frac{\rho}{2}\|\xb_i -\xb_0^{k+1}\|^2, \notag \\
				&~~~~~~~~~~~~~~~~~~~~~~~~~~~~~~~~\forall\; i=1,\ldots,N,
				\label{eqn: sync cadmm s1}\\
				\lambdab^{k+1}_i&= \lambdab^{k}_i + \rho (\xb_i^{k+1}-\xb_0^{k+1}),~\forall\; i=1,\ldots,N.\label{eqn: sync cadmm s3}
			\end{align}
			
			\STATE {\bf set} $k\leftarrow k+1.$
			\UNTIL {a predefined stopping criterion is satisfied.}%
		\end{algorithmic}
	\end{algorithm}

	As seen, Algorithm \ref{table: sync cadmm} is naturally implementable over the star computer network illustrated in Figure \ref{fig: masterslave network}. Specifically, the master node takes charge of optimizing $\xb_0$ by \eqref{eqn: sync cadmm s2}, and each worker $i$ is responsible for optimizing $(\xb_i, \lambdab_i)$ by \eqref{eqn: sync cadmm s1}-\eqref{eqn: sync cadmm s3}.
	Through exchanging the up-to-date $\xb_0$ and $(\xb_i,\lambdab_i)$ between the master and the workers, Algorithm \ref{table: sync cadmm} solves problem \eqref{eqn: original problem} in a fully distributed and parallel manner. Convergence properties of the distributed ADMM have been extensively studied; see, e.g., \cite{BoydADMM11,He12ADMMite,DengYin2013J,HongLuo2013,Hong15noncvxadmm}.
	Specifically, \cite{He12ADMMite} shows that the ADMM, under general convex assumptions, has a worst-case $\mathcal{O}(1/k)$ convergence rate; while \cite{DengYin2013J} shows that the ADMM can have a linear convergence rate given strong convexity and smoothness conditions on $f_i$'s.
	For non-convex and smooth $f_i$'s, the work \cite{Hong15noncvxadmm} shows that Algorithm \ref{table: sync cadmm} can converge to the set of KKT points with a $\mathcal{O}(1/\sqrt{k})$ rate as long as $\rho$ is large enough.

	However, Algorithm \ref{table: sync cadmm} is a synchronous algorithm, where the operations of the master and the workers are ``locked" with each other. Specifically, to optimize $\xb_0$ at each iteration, the master has to wait until receiving all the up-to-date variables $(\xb_i,\lambdab_i)$, $i=1,\ldots,N$, from the workers. Since the workers may have different computation and communication delays\footnote{In a heterogeneous network, the workers can have different computational powers, or the data sets can be non-uniformly distributed across the network.
		Thus, the workers can require different computational times in solving the local subproblems. Besides, the communication delays can also be different, e.g., due to probabilistic communication failures and message retransmission.}, the pace of the optimization would be determined by the ``slowest" worker.
	As an example illustrated in Figure \ref{fig: sync vs async}(a), the master updates $\xb_0$ only when it has received the variable information for the four workers  at every iteration. As a result, under such synchronous protocol, the master and speedy workers (e.g., workers 1 and 3 in Figure \ref{fig: sync vs async}) would spend most of the time idling, and thus the parallel computational resources cannot be fully utilized.

	\section{Asynchronous Distributed ADMM}\label{sec: async D-ADMM}
	
	\subsection{Algorithm Description}
\begin{figure}[h]
		\begin{center}
			{\subfigure[][Synchronous distributed ADMM]{\resizebox{.45\textwidth}{!}
					{\includegraphics{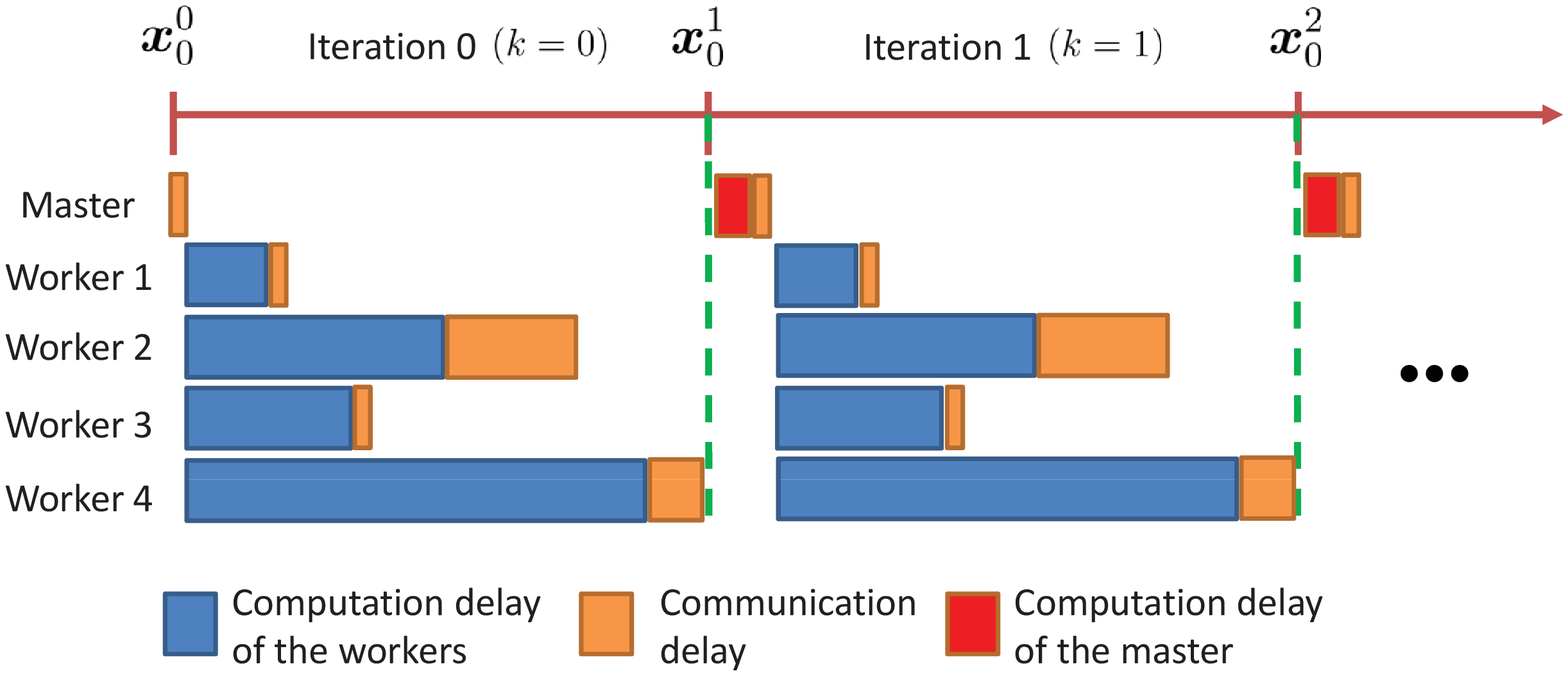}}}
			}
			\hspace{0pc}
			{\subfigure[][Asynchronous distributed ADMM]{\resizebox{.45\textwidth}{!}{\includegraphics{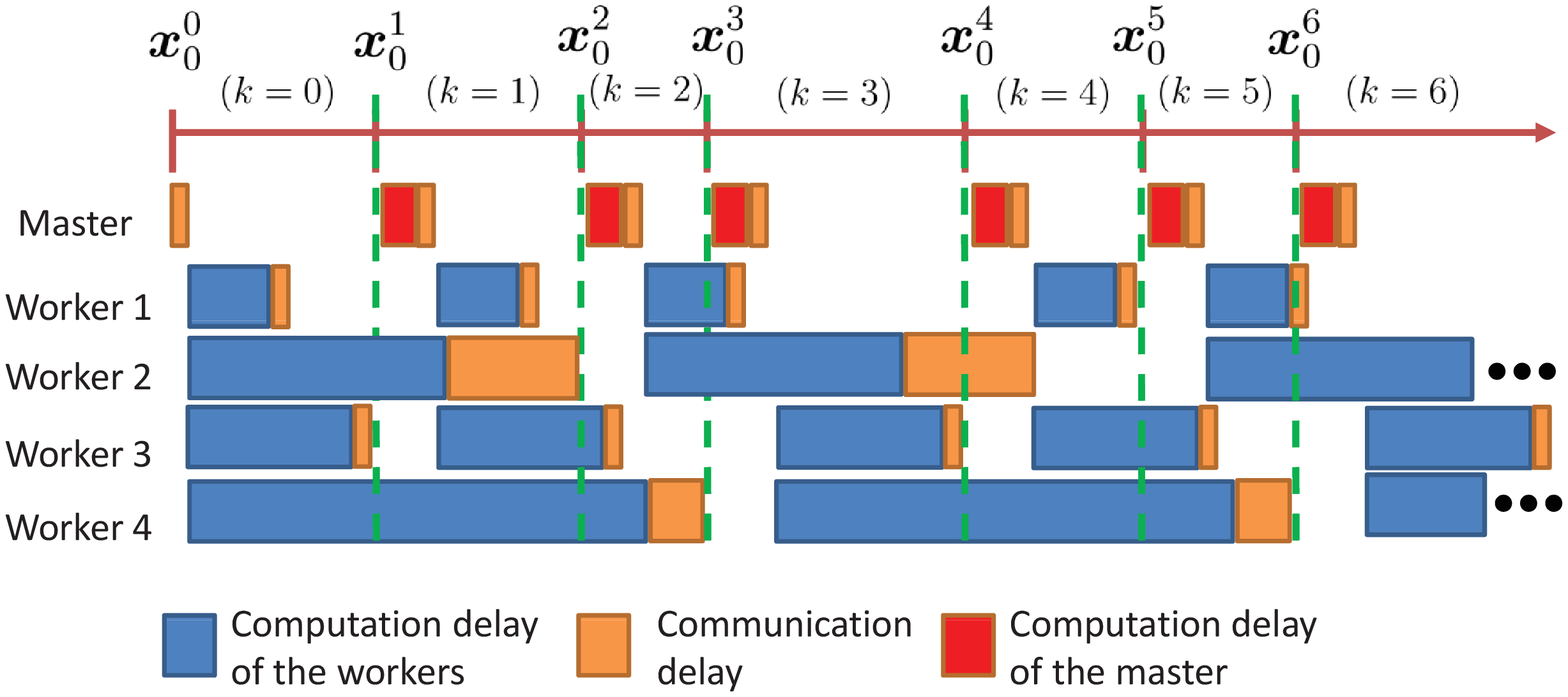}}}
			}
		\end{center}\vspace{-0.2cm}
		\caption{Illustration of synchronous and asynchronous distributed ADMM.}
		\vspace{-0.5cm}\label{fig: sync vs async}
	\end{figure}

	In this section, we present an AD-ADMM. The asynchronism we consider is in the same spirit of \cite{NiuNIPS11,AgarwalNIPS11,MLiPS14,MLiNIPS14,LiuWright15,Zhang14ACADMM,Hong15noncvxadmm2}, where the master
	does not wait for all the workers. Instead, the master updates $\xb_0$ whenever it receives $(\xb_i,\lambdab_i)$ from a partial set of the workers. For example, in Figure \ref{fig: sync vs async}(b), the master updates $\xb_0$ whenever it receives the variable information from at least two workers.
	This implies that {none of the workers have to be synchronized with each other and the master does not need to wait for the slowest worker either}. As illustrated in Figure \ref{fig: sync vs async}(b), with the lock removed, both the master and speedy workers can update their variables more frequently.
	
	Let us denote {$k\geq 0$} as the iteration number of the master (i.e., the number of times for which the master updates $\xb_0$), and denote $\Ac_k \subseteq \Vc\triangleq \{1,\ldots,N\}$ as the index subset of workers from which the master receives variable information during iteration $k$ (for example, in Figure \ref{fig: sync vs async}(b), $\Ac_0=\{1,3\}$ and  $\Ac_1=\{1,2\}$)\footnote{Without loss of generality, we let $\Ac_{-1}=\Vc$, as seen from Figure \ref{fig: sync vs async}.}. We say that worker $i$ is ``arrived" at iteration $k$ if $i\in \Ac_k$ and ``unarrived" otherwise.
	{
		Clearly, unbounded delay will jeopardize the algorithm convergence. Therefore throughout this paper,  we will assume that the asynchronous delay in the network is bounded. In particular, we follow the popular {\it partially asynchronous} model \cite{BertsekasADMM} and assume: }		
	\begin{Assumption}\label{assumption bounded delay} {\rm (Bounded delay)}
		Let $\tau\geq 1$ be a maximum tolerable delay. For all $i\in \Vc$ and iteration {$k\geq 0$,} it must be that $ i\in \Ac_k \cup \Ac_{k-1}\cdots \cup \Ac_{\max\{k-\tau+1,-1\}}$.
	\end{Assumption}
	
	Assumption \ref{assumption bounded delay} implies that every worker $i$ is arrived at least once within the period $[k-\tau+1, k]$. In another word, the variable information $(\xb_i,\lambdab_i)$ used by the master must be at most $\tau$ iterations old. To guarantee the bounded delay, at every iteration the master should wait for the workers who have been inactive for $\tau-1$ iterations, if such workers exist. 
	Note that, when $\tau=1$, one has $i\in \Ac_k$ for all $i\in \Vc$ (i.e., $\Ac_k=\Vc$), which corresponds to the synchronous case and the master always waits for all the workers at every iteration.

	In Algorithm \ref{table: async cadmm s1 master}, we present the proposed AD-ADMM, which specifies respectively the steps for the master and the distributed workers. 
	Here, $\Ac_k^c$ denotes the complementary set of $\Ac_k$, i.e., $\Ac_k\cap \Ac_k^c =\emptyset$ and $\Ac_k\cup \Ac_k^c=\Vc$. Algorithm \ref{table: async cadmm s1 master} has {five} notable differences compared with Algorithm \ref{table: sync cadmm}. First, the master is required to update $\{(\xb_i,\lambdab_i)\}_{i\in \Vc}$, and such update is only performed for those variables with $i\in \Ac_k$. Second, $\xb_0$ is updated by solving a problem with an additional proximal term $\frac{\gamma}{2}\|\xb_0-\xb_0^k\|^2$, where $\gamma>0$ is a penalty parameter (cf. \eqref{eqn: async cadmm s1 x0}). Adding such proximal term is crucial in making the algorithm well-behaved in the asynchronous setting. As will be seen in the next section, a proper choice of $\gamma$ guarantees the convergence of Algorithm \ref{table: async cadmm s1 master}. Third, the variables $d_i$'s are introduced to count the delays of the workers. If worker $i$ is arrived at the current iteration, then $d_i$ is set to zero; otherwise, $d_i$ is increased by one.
	{So, to ensure Assumption \ref{assumption bounded delay} hold all the time, in Step 4 of \emph{Algorithm of the Master}, the master waits if there exists at least one worker whose $d_i\geq \tau-1$. Fourth, in addition to the bounded delay, we assume that the master proceeds to update the variables only if there are at least $A\geq 1$ {arrived} workers, i.e., $|\Ac_k|\geq A$ for all $k$ \cite{Zhang14ACADMM}. Note that when $A=N$, the algorithm reduces to the synchronous {distributed} ADMM.}
	Fifth, in Step 6 of \emph{Algorithm of the Master}, the master sends the up-to-date $\xb_0$ only to the arrived workers.
	
	{We emphasize again that both the master and fast workers in the AD-ADMM {can have less idle time and update more frequently} than its synchronous counterpart.}  As illustrated in Figure \ref{fig: sync vs async}, during the same period of time, the synchronous algorithm only completes two updates whereas the asynchronous algorithm updates six times already. On the flip side, the asynchronous algorithm introduces delayed variable information and thereby requires a larger number of iterations to reach the same solution accuracy {than} its synchronous counterpart.
	{In practice we observe that the benefit of improved update frequency can outweigh the cost of increased number of iterations, and as a result the asynchronous algorithm can still converge faster in time. This is particularly true when the workers have different computation and communication delays and when	
		the computation and communication delays of the master for solving \eqref{eqn: async cadmm s1 x0} is much shorter than the computation and communication delays of the workers for updating \eqref{eqn: async cadmm s1 xi} and \eqref{eqn: async cadmm s1 lambda}\footnote{
			Note that, for many {practical} cases (such as $h(\xb_0)=\|\xb_0\|_1$) for which \eqref{eqn: async cadmm s1 x0} has a {closed-form} solution,  the {computation} delay of the master is {negligible}. For high-performance computer clusters connected by large-bandwidth fiber links, the communication delays between the master and the workers can also be short. However, for cases in which the computation and communication delays of the master is significant, the AD-ADMM could be less time efficient than the synchronous ADMM due to the increased number of iterations.}; e.g., see Figure \ref{fig: sync vs async}.}
	Detailed numerical results will be reported in Section V of the companion paper \cite{ChangAsyncadmm15_p2}.

	\begin{algorithm}[h!] \label{table: async cadmm s1}
		\caption{Asynchronous Distributed ADMM for \eqref{eqn: consensus problem}.}
		\begin{algorithmic}[1]\label{table: async cadmm s1 master}
			\STATE {\bf \underline{Algorithm of the Master:}}
			\STATE {\bf Given} initial variable
			$\xb^{0}$ and broadcast it to the workers. Set $k=0$ and $d_1=\cdots=d_N=0;$
			\REPEAT
			\STATE  {\bf wait} {until receiving $\{\hat\xb_i,\hat\lambdab_i\}_{i\in \Ac_k}$ from workers $i\in \Ac_k$ such that  $|\Ac_k|\geq A$ and $d_i <\tau-1$ $\forall i \in \Ac_k^c$.}
			\STATE {\bf update}
			\begin{align}&\xb^{k+1}_i =\bigg\{\begin{array}{ll}
			\hat\xb_i  & \forall i\in \Ac_k \\
			\xb^{k}_i & \forall i\in \Ac_k^c
			\end{array},  \label{eqn: async cadmm s1 xii} \\
			&\lambdab^{k+1}_i =\bigg\{\begin{array}{ll}
			\hat\lambdab_i  & \forall i\in \Ac_k \\
			\lambdab^{k}_i & \forall i\in \Ac_k^c
			\end{array}, \label{eqn: async cadmm s1 lambdaii} \\
			&d_i =\bigg\{\begin{array}{ll}
			0  & \forall i\in \Ac_k \\
			d_i+1 & \forall i\in \Ac_k^c
			\end{array}, \\
			&\xb^{k+1}_0 \! =\!\arg\min_{\xb_0 \in \mathbb{R}^n} \textstyle  \bigg\{h(\xb_0)-  \xb_0^T\sum_{i=1}^N \lambdab_i^{k+1} \notag \\
			 &~\textstyle+\frac{\rho}{2}\sum_{i=1}^N\|\xb_i^{k+1}-\xb_0\|^2+\frac{\gamma}{2}\|\xb_0-\xb_0^k\|^2\bigg\},\label{eqn: async cadmm s1 x0}
			\end{align}
			\STATE {\bf broadcast} $\xb^{k+1}_0$ to the workers in $\Ac_k$.
			\STATE {\bf set} $k\leftarrow k+1.$
			\UNTIL {a predefined stopping criterion is satisfied.}%
		\end{algorithmic}
		
		\begin{algorithmic}[1]\label{table: async cadmm s1 worker}
			
			\STATE {\bf \underline{Algorithm of the $i$th Worker:}}
			\STATE {\bf Given} initial $\lambdab^{0}$ and set ${k_i}=0.$
			\REPEAT
			\STATE  {\bf wait} until receiving $\hat\xb_0$ from the master node.
			\STATE {\bf update}
			\begin{align}
			\xb_i^{{k_i}+1} &=\arg\min_{\xb_i\in \mathbb{R}^n} \textstyle f_i(\xb_i)+\xb_i^T\lambdab_i^{k_i}+\frac{\rho}{2}\|\xb_i -\hat\xb_0\|^2,
			\label{eqn: async cadmm s1 xi}\\
			\lambdab_i^{{k_i}+1}&= \lambdab^{{k_i}}_i + \rho (\xb_i^{{k_i}+1}-\hat \xb_0).\label{eqn: async cadmm s1 lambda}
			\end{align}
			\STATE {\bf send} $(\xb_i^{{k_i}+1},\lambdab_i^{{k_i}+1})$ to the master node.
			\STATE {\bf set} ${k_i}\leftarrow {k_i}+1.$
			\UNTIL {a predefined stopping criterion is satisfied.}%
		\end{algorithmic}
	\end{algorithm}

	
	%
	
	\subsection{Convergence Analysis}
	
	In this subsection, we analyze the convergence conditions of Algorithm \ref{table: async cadmm s1 master}.
	We first make the following standard assumption on problem \eqref{eqn: original problem} (or equivalently problem \eqref{eqn: consensus problem}):
	\begin{Assumption} \label{assumption obj s1}
		Each function $f_i$ is twice differentiable and its gradient $\nabla f_i$ is Lipschitz continuous with a Lipschitz constant $L>0$; {the function $h$ is proper convex (lower semi-continuous, but not necessarily smooth) and dom$(h)$ (the domain of $h$) is compact.} 
		Moreover, problem \eqref{eqn: original problem} is bounded below, i.e., $F^\star >-\infty$ where $F^\star$ denotes the optimal objective value of problem \eqref{eqn: original problem}.
	\end{Assumption}
	Notably, we do not assume any convexity on $f_i$'s. Indeed, we will show that the AD-ADMM can converge to the set of KKT points even for non-convex $f_i$'s. Our main result is formally stated below.
	\begin{Theorem} \label{thm: conv of scheme 1}Suppose that Assumption \ref{assumption bounded delay} and Assumption \ref{assumption obj s1} hold true. Moreover, assume that there exists a constant $S\in [1, N]$ such that $|\Ac_k|<S$ for all $k$ and that 
		\begin{align}
			&{\infty > \Lc_\rho(\xb^0,\xb_0^0,\lambdab^0) - F^\star \geq 0, \label{eqn: cond of conv scheme1 L}}\\
			&\rho > \frac{(1+L+L^2)+\sqrt{(1+L+L^2)^2+8L^2}}{2}, \label{eqn: cond of conv scheme1 rho} \\
			&\gamma > \frac{S(1+\rho^2)(\tau-1)^2 -N\rho}{2}. \label{eqn: cond of conv scheme1 gamma}
		\end{align}
		{Then,  $(\{\xb_i^k\}_{i=1}^N,\xb_0^k, \{\lambdab_i^k\}_{i=1}^N)$ generated by \eqref{eqn: async cadmm s1 xii}, \eqref{eqn: async cadmm s1 lambdaii} and \eqref{eqn: async cadmm s1 x0} are bounded and have limit points which satisfy
         KKT conditions of problem \eqref{eqn: consensus problem}.}
		
	\end{Theorem}
	Theorem \ref{thm: conv of scheme 1} implies that the AD-ADMM is guaranteed to converge to the set of KKT points as long as the penalty parameters $\rho$ and $\gamma$ are sufficiently large.
	{Since $1/\gamma$ can be viewed as the step size of $\xb_0$,} \eqref{eqn: cond of conv scheme1 gamma} indicates that the master should be more cautious in moving $\xb_0$ if the network allows a longer delay $\tau$.
	In particular, the value $\gamma$ in the worst case should increase with the order of $\tau^2$.
	When $\tau=1$ (the synchronous case), $\gamma=-(N\rho)/2<0$ and thus the proximal term $\frac{\gamma}{2}\|\xb_0-\xb_0^k\|^2$ can be removed from \eqref{eqn: async cadmm s1 x0}.  {On the other hand, we also see from \eqref{eqn: cond of conv scheme1 gamma} that $\gamma$ should increase with $N$ if $\tau>1$ is fixed\footnote{Note that, for a fixed $\tau$, $S$ should increase with $N$.}.
    This is because in the worst case the more workers, the more {outdated} information introduced in the network.
	Finally, we should mention that a large $\rho$ may be essential for the AD-ADMM to converge properly, especially for non-convex problems, as we demonstrate via simulations in Section \ref{sec: simu}.}

	Let us compare Theorem \ref{thm: conv of scheme 1} with the results in \cite{Zhang14ACADMM,ErminWei2013arxiv}. First, the convergence conditions in \cite{Zhang14ACADMM,ErminWei2013arxiv} are only applicable for convex problems, whereas our results hold for both convex and non-convex problems. Second, \cite{Zhang14ACADMM,ErminWei2013arxiv} have made specific statistical assumptions on the behavior of the workers, and the convergence results presented therein are in an expectation sense. Therefore it is possible, at least theoretically, that a realization of the algorithm fails to converge despite satisfying the conditions given in \cite{Zhang14ACADMM}. On the contrary, our convergence results hold deterministically. 

	Note that for non-convex $f_i$'s, subproblem \eqref{eqn: async cadmm s1 xi} is not necessarily convex. However, given $\rho\geq L$ in \eqref{eqn: cond of conv scheme1 rho} and {twice {differentiability} of $f_i$ (Assumption \ref{assumption obj s1}),} 
	subproblem \eqref{eqn: async cadmm s1 xi} becomes a (strongly) convex problem\footnote{By \cite[Lemma 1.2.2]{BK:Nesterov05}, the minimum eigenvalue of the Hessian matrix of $f_i(\xb_i)$ is no smaller than $-L$.
		Thus, for $\rho>L$, subproblem \eqref{eqn: async cadmm s1 xi} is a strongly convex problem.} and hence is globally solvable. When $f_i$'s are all convex functions, Theorem \ref{thm: conv of scheme 1} reduces to the following corollary.
	
	\begin{Corollary}\label{corollary: conv of scheme 1 convex} Assume that $f_i$'s are all convex functions. Under the same premises of Theorem
		\ref{thm: conv of scheme 1}, and for $\gamma$ satisfying \eqref{eqn: cond of conv scheme1 gamma} and
		\begin{align}
			&\rho \geq \frac{(1+L^2)+\sqrt{(1+L^2)^2+8L^2}}{2}, \label{eqn: cond of conv scheme1 rho2}
		\end{align}
		{ $(\{\xb_i^k\}_{i=1}^N,\xb_0^k, \{\lambdab_i^k\}_{i=1}^N)$ generated by \eqref{eqn: async cadmm s1 xii}, \eqref{eqn: async cadmm s1 lambdaii} and \eqref{eqn: async cadmm s1 x0} are bounded and have limit points which satisfy
			KKT conditions of problem \eqref{eqn: consensus problem}. }
	\end{Corollary}
	

	\subsection{Proof of Theorem \ref{thm: conv of scheme 1} and Corollary \ref{corollary: conv of scheme 1 convex}}\label{subsec: proof of thm1}
	
	Let us write Algorithm \ref{table: async cadmm s1 master} from the master's point of view.
	Define $\bar k_i$ as the last iteration number before iteration $k$ for which worker $i\in \Ac_k$ is {arrived\footnote{Note that
		$\bar k_i=-1$ for $k=0$ and  $\bar k_i\geq -1$ for $k\geq 0$},} i.e., $i\in \Ac_{\bar k_i}$. Then Algorithm \ref{table: async cadmm s1 master} from the master's point of view is
	as follows: for master iteration $k=0,1,\ldots,$
	\begin{align}
		\xb^{k+1}_i &\!=\!\left\{\!\!\begin{array}{ll}
			\arg{\displaystyle \min_{\xb_i}} \textstyle \bigg\{f_i(\xb_i)+\xb_i^T\lambdab^{\bar k_i +1}_i & \\
			~~~~~~~~~~~~+\frac{\rho}{2}\|\xb_i -\xb_0^{\bar k_i +1}\|^2\bigg\}, & \forall i\in \Ac_k \\
			\xb^{k}_i & \forall i\in \Ac_k^c
		\end{array}\right.,\label{eqn: async cadmm s1 xi equi0}
		\\
		\lambdab^{k+1}_i &=\bigg\{\begin{array}{ll}
			\lambdab^{\bar k_i +1}_i + \rho (\xb_i^{k+1}-\xb_0^{\bar k_i +1})  & \forall i\in \Ac_k \\
			\lambdab^{k}_i & \forall i\in \Ac_k^c
		\end{array}, \label{eqn: async cadmm s1 lambda equi0} \\
		\xb^{k+1}_0 \! &=\!\arg\min_{\xb_0 \in \mathbb{R}^n} \textstyle  \bigg\{h(\xb_0)-  \xb_0^T\sum_{i=1}^N \lambdab_i^{k+1} \notag \\
		 &~~~~~~~~~\textstyle+\frac{\rho}{2}\sum_{i=1}^N\|\xb_i^{k+1}-\xb_0\|^2+\frac{\gamma}{2}\|\xb_0-\xb_0^k\|^2\bigg\}.\notag
	\end{align}
	Now it is relatively easy to see that the master updates $\xb_0$ using the delayed $(\xb_i,\lambdab_i)_{i\in \Ac_k}$ and the old $(\xb_i,\lambdab_i)_{i\in \Ac_k^c}$.
	Under Assumption 1, it must hold
	\begin{align}\label{eqn: cond on bar ki}
		 \max\{k-\tau, -1\} \leq  \bar k_i <k~~\forall \; k\geq 0.
	\end{align}
	Moreover, by the definition of $\bar{k}_i$ it holds that $i\notin \Ac_{k-1}\cup \cdots \cup \Ac_{\bar k_i+1}$, therefore we have that
	\begin{align}\label{eqn: unchanged lambda}
		\lambdab^{\bar k_i +1}_i = \lambdab^{\bar k_i +2}_i =\cdots=\lambdab^{k}_i,\quad \forall\; i\in \Ac_k.
	\end{align} By applying \eqref{eqn: unchanged lambda} to \eqref{eqn: async cadmm s1 xi equi0} and \eqref{eqn: async cadmm s1 lambda equi0} (replacing $\lambdab^{\bar k_i+1}_i$  with $\lambdab^k_i$), we rewrite the master-point-of-view algorithm in Algorithm \ref{table: async cadmm global view}. 
	
	\begin{algorithm}[h!]
		\caption{Asynchronous distributed ADMM from the master's point of view.}
		\begin{algorithmic}[1]\label{table: async cadmm global view}
			\STATE {\bf Given} initial variables
			$\xb^{0}$ and $\lambdab^{0}$; set $\xb_0^0=\xb^{0}$ and $k=0.$
			\REPEAT
			\STATE  {\bf update}
			\begin{align}
				\xb^{k+1}_i &\!=\!\left\{\!\!\begin{array}{ll}
					\arg{\displaystyle \min_{\xb_i\in \mathbb{R}^n}} \textstyle \bigg\{f_i(\xb_i)+\xb_i^T\lambdab^{k}_i & \\
					~~~~~~~~~~~~+\frac{\rho}{2}\|\xb_i -\xb_0^{\bar k_i +1}\|^2\bigg\}, & \forall i\in \Ac_k \\
					\xb^{k}_i & \forall i\in \Ac_k^c
				\end{array}\right.,\label{eqn: async cadmm s1 xi equi} \\
				\lambdab^{k+1}_i &=\bigg\{\begin{array}{ll}
					\lambdab^{k}_i + \rho (\xb_i^{k+1}-\xb_0^{\bar k_i +1})  & \forall i\in \Ac_k \\
					\lambdab^{k}_i & \forall i\in \Ac_k^c
				\end{array}, \label{eqn: async cadmm s1 lambda equi} \\
				\xb^{k+1}_0 \! &=\!\arg\min_{\xb_0 \in \mathbb{R}^n} \textstyle  \bigg\{h(\xb_0)-  \xb_0^T\sum_{i=1}^N \lambdab_i^{k+1} \notag \\
				 &~\textstyle+\frac{\rho}{2}\sum_{i=1}^N\|\xb_i^{k+1}-\xb_0\|^2+\frac{\gamma}{2}\|\xb_0-\xb_0^k\|^2\bigg\}.\label{eqn: async cadmm s1 x0 equi}
			\end{align}
			
			\STATE {\bf set} $k\leftarrow k+1.$
			\UNTIL {a predefined stopping criterion is satisfied.}%
		\end{algorithmic}
	\end{algorithm}

	Inspired by \cite{Hong15noncvxadmm}, our analysis for Theorem \ref{thm: conv of scheme 1} investigates how the augmented Lagrangian function, i.e.,
	\begin{align}\label{eqn: Lc}
		\Lc_{\rho}(\xb^k,\xb_0^k, \lambdab^k)=
		\sum_{i=1}^N f_i(\xb_i^k) &+ h(\xb_0^k)+\sum_{i=1}^N (\lambdab_i^k)^T(\xb_i^k -\xb_0^k) \notag \\
		&+\frac{\rho}{2}\sum_{i=1}^N\|\xb_i^k -\xb_0^k\|^2
	\end{align} evolves with the iteration number $k$, where $\xb^k\triangleq[(\xb^k_1)^T,\ldots,(\xb^k_N)^T]^T$ and $\lambdab^k\triangleq[(\lambdab^k_1)^T,\ldots,(\lambdab^k_N)^T]^T$.
	The following lemma is one of the keys to prove Theorem \ref{thm: conv of scheme 1}.
	
	\begin{Lemma} \label{lemma: Lc progress} Suppose that Assumption \ref{assumption obj s1} holds and $\rho\geq L$. Then, 
		{it holds that}
		\begin{align}
			&\Lc_{\rho}(\xb^{k+1},\xb_0^{k+1}, \lambdab^{k+1}) -
			\Lc_{\rho}(\xb^k,\xb_0^k, \lambdab^k) \notag \\
			&\leq -\frac{2\gamma+N\rho}{2}\|\xb_0^{k+1}-\xb_0^k\|^2 \notag \\
			&~~~+ \bigg(\frac{1}{\rho}+\frac{1}{2}\bigg)\sum_{i\in \Ac_k} \|\lambdab_i^{k+1}-\lambdab_i^k\|^2
			\notag \\
			&~~~+\frac{1+\rho^2}{2}\!\sum_{i\in \Ac_k}\! \|\xb_0^{\bar k_i+1}\!-\!\xb_0^{k}\|^2\! \notag \\
			&~~~+\!\frac{(1-\rho)+L}{2}\!\sum_{i\in \Ac_k}\!\|\xb_i^{k+1}\! -\!\xb_i^{k}\|^2.\label{lemma: Lc progress eq 7}
		\end{align}
	\end{Lemma}
	{\bf Proof:} See Appendix \ref{appx: proof of lemma: Lc progress}. \hfill $\blacksquare$
	
	Equation \eqref{lemma: Lc progress eq 7} shows that $\Lc_{\rho}(\xb^k,\xb_0^k, \lambdab^k)$ is not necessarily decreasing due to the error terms $\sum_{i\in \Ac_k} \|\lambdab_i^{k+1}-\lambdab_i^k\|^2$ and $\sum_{i\in \Ac_k}\! \|\xb_0^{\bar k_i+1}\!-\!\xb_0^{k}\|^2$. Next we bound the sizes of these two terms.
	
	First consider $\sum_{i\in \Ac_k} \|\lambdab_i^{k+1}-\lambdab_i^k\|^2$.
	Note from \eqref{eqn: async cadmm s1 lambda equi} and
	the optimality condition of \eqref{eqn: async cadmm s1 xi equi} that, $\forall\; i\in \Ac_k$,
	\begin{align}
		\zerob&=\nabla f_i(\xb_i^{k+1}) + \lambdab^{k}_i + \rho(\xb_i^{k+1} -\xb_0^{\bar k_i+1})
		\notag \\
		&=\nabla f_i(\xb_i^{k+1})+\lambdab_i^{k+1}. \label{lemma: Lc progress eq 5}
	\end{align}
	For any $i\in \Ac_k^c$, denote $\widetilde k_i<k$ as the last iteration number for which
	worker $i$ is arrived. Then, $i \in \Ac_{\widetilde k_i}$ and thus $\nabla f_i(\xb_i^{\widetilde k_i+1})+\lambdab_i^{\widetilde k_i+1}=\zerob$. Since $\xb_i^{\widetilde k_i+1}=\xb_i^{\widetilde k_i+2}=\cdots=\xb_i^{k}=\xb_i^{k+1}$ and $\lambdab_i^{\widetilde k_i+1}=\lambdab_i^{\widetilde k_i+2}=\cdots=\lambdab_i^{k}=\lambdab_i^{k+1}$, we obtain that $\nabla f_i(\xb_i^{k+1})+\lambdab_i^{k+1}=\zerob\; \forall i\in \Ac_k^c$. Therefore, we conclude that
	\begin{align}\label{lemma: Lc progress eq 8}
		\nabla f_i(\xb_i^{k+1})+\lambdab_i^{k+1}=\zerob, \quad\forall~i\in \Vc ~{\rm and}~ \forall~k.
	\end{align}
	By \eqref{lemma: Lc progress eq 8} and the Lipschitz continuity of $\nabla f_i$ (Assumption \ref{assumption obj s1}), we can bound
	\begin{align}\label{lemma: Lc progress eq 9}
		\|\lambdab_i^{k+1}-\lambdab_i^k\|^2 &\leq \|\nabla f_i(\xb_i^{k+1})-\nabla f_i(\xb_i^{k})\|^2 \notag \\
		&\leq L^2 \|\xb_i^{k+1}-\xb_i^{k}\|^2, \quad \forall~i\in \Vc.
	\end{align}
	By applying \eqref{lemma: Lc progress eq 9}, we can further write \eqref{lemma: Lc progress eq 7} as
	\begin{align}
		&\Lc_{\rho}(\xb^{k+1},\xb_0^{k+1}, \lambdab^{k+1}) \notag \\
		&~~~~ \leq
		\Lc_{\rho}(\xb^k,\xb_0^k, \lambdab^k)+ \bigg(\frac{1+\rho^2}{2}\bigg)\sum_{i\in \Ac_k} \|\xb_0^{k}-\xb_0^{\bar k_i+1}\|^2
		\notag \\
		&~~~~-
		\bigg(\frac{2\gamma+N\rho}{2}\bigg)\|\xb_0^{k+1}-\xb_0^{k}\|^2
		\notag \\
		&~~~~+ \bigg(\frac{L+L^2+(1-\rho)}{2}+\frac{L^2}{\rho}\bigg)\sum_{i\in \Ac_k} \|\xb_i^{k+1}-\xb_i^{k}\|^2. \label{lemma: Lc progress eqn0}
	\end{align}
	
	From \eqref{lemma: Lc progress eqn0}, one can observe that the error term
	$(\frac{1+\rho^2}{2})\sum_{i\in \Ac_k} \|\xb_0^{k}-\xb_0^{\bar k_i+1}\|^2$ is present due to the asynchrony of the network.
	The next lemma bounds this error term:
	\begin{Lemma}\label{lemma: asyn error bound}
		Suppose that Assumption \ref{assumption bounded delay} holds and assume that $|\Ac_k|<S$ for all $k$, for some constant $S\in [1,N]$. Then, it holds that
		\begin{align}
			\sum_{j=0}^{k}\sum_{i\in \Ac_j} \|\xb_0^{j}-\xb_0^{\bar j_i+1}\|^2
			&\leq S(\tau-1)^2 \sum_{j=0}^{k-1}\|\xb_0^{j+1}-\xb_0^{j}\|^2. \label{lemma: asyn error bound eq0}
		\end{align}
	\end{Lemma}
	{\bf Proof:} See Appendix \ref{appx: proof of lemma: asyn error bound}. \hfill $\blacksquare$
	
	The last lemma  shows that $\Lc_{\rho}(\xb^k,\xb_0^k, \lambdab^k)$ is bounded below:
	\begin{Lemma} \label{lemma: Lc lower bounded}
		Under Assumption \ref{assumption obj s1} and for $\rho \geq L$, it holds that
		\begin{align}\label{eqn: Lc lower bounded}
			\Lc_{\rho}(\xb^{k+1},\xb_0^{k+1}, \lambdab^{k+1}) 
			&\geq F^\star >-\infty.
		\end{align}
	\end{Lemma}
	{\bf Proof:} See Appendix \ref{appx: proof of lemma: Lc lower bounded}. \hfill $\blacksquare$
	
	Given the three lemmas above, we are ready to prove Theorem \ref{thm: conv of scheme 1}. 
	
	{\bf Proof of Theorem \ref{thm: conv of scheme 1}:}
	Note that any KKT point $(\{\xb_i^\star\}_{i=1}^N,\xb_0^\star,\{\lambdab_i^\star\}_{i=1}^N)$ of problem \eqref{eqn: consensus problem} satisfies the following conditions
	\begin{subequations}\label{eqn: KKT of prob}
		\begin{align}
			&\nabla f_i(\xb_i^\star) +\lambdab_i^\star =\zerob, \quad\forall~i\in \Vc, \\
			&{\ssb_0^\star - \textstyle \sum_{i=1}^N \lambdab_i^\star = \zerob, }\\
			& \xb_i^\star =\xb_0^\star, \quad\forall~i\in \Vc,
		\end{align}
	\end{subequations}
	{where $\ssb_0^\star \in \partial h(\xb_0^\star)$ denotes a subgradient of $h$
		at $\xb_0^\star$ and $\partial h(\xb_0^\star)$ is the subdifferential of $h$
		at $\xb_0^\star$.} Since \eqref{eqn: KKT of prob} also implies
	\begin{align}
		&\sum_{i=1}^N\nabla f_i(\xb^\star) {+\ssb_0^\star}=\zerob,
	\end{align} where $\xb^\star \triangleq \xb_0^\star = \cdots=\xb_N^\star,$ $\xb^\star$ is also a stationary point of the original problem \eqref{eqn: original problem}.

	To prove the desired result, we take a telescoping sum of \eqref{lemma: Lc progress eqn0}, which yields
	\begin{align}
		&\Lc_{\rho}(\xb^{k+1},\xb_0^{k+1}, \lambdab^{k+1})-\Lc_{\rho}(\xb^0,\xb_0^0, \lambdab^0) \notag \\
		&\leq
		\bigg(\frac{1+\rho^2}{2}\bigg)\sum_{j=0}^{k}\sum_{i\in \Ac_j} \|\xb_0^{j}-\xb_0^{\bar j_i+1}\|^2
		\notag \\
		&~+ \bigg(\frac{L+L^2+(1-\rho)}{2}+\frac{L^2}{\rho}\bigg)\sum_{j=0}^{k}\sum_{i\in \Ac_j} \|\xb_i^{j+1}-\xb_i^{j}\|^2 \notag \\
		&~-
		\bigg(\frac{2\gamma+N\rho}{2}\bigg)\sum_{j=0}^{k}\|\xb_0^{j+1}-\xb_0^{j}\|^2. \label{thm: Lc progress eqn1}
	\end{align}
	By substituting \eqref{lemma: asyn error bound eq0} in Lemma \ref{lemma: asyn error bound}
	into \eqref{thm: Lc progress eqn1}, we obtain
	\begin{align}
		&\bigg(\frac{2\gamma+N\rho-S(1+\rho^2)(\tau-1)^2}{2}\bigg)\sum_{j=0}^{k-1}\|\xb_0^{j+1}-\xb_0^{j}\|^2
		\notag \\
		&+\bigg(\frac{(1-\rho)-(L+L^2)}{2}-\frac{L^2}{\rho}\bigg)\sum_{j=0}^{k}\sum_{i=1}^N \|\xb_i^{j+1}-\xb_i^{j}\|^2 \notag \\
		&\leq \Lc_{\rho}(\xb^0,\xb_0^0, \lambdab^0) -\Lc_{\rho}(\xb^{k+1},\xb_0^{k+1}, \lambdab^{k+1}) \notag \\
		& = (\Lc_{\rho}(\xb^0,\xb_0^0, \lambdab^0) -F^\star) - (\Lc_{\rho}(\xb^{k+1},\xb_0^{k+1}, \lambdab^{k+1})-F^\star)
		\notag \\
		&\leq \Lc_{\rho}(\xb^0,\xb_0^0, \lambdab^0) -F^\star <\infty,
		\label{thm: Lc progress eqn4}
	\end{align}
	where the second inequality is obtained by applying Lemma \ref{lemma: Lc lower bounded}, and the last strict inequality is due to Assumption \ref{assumption obj s1} where the optimal value $F^\star$ is assumed to be lower bounded.
	
	Then, \eqref{eqn: cond of conv scheme1 rho} and \eqref{eqn: cond of conv scheme1 gamma} imply that the left hand side (LHS) of \eqref{thm: Lc progress eqn4} is positive and increasing with $k$.
	Since the RHS of \eqref{thm: Lc progress eqn4} is finite, we must have, as $k\to \infty,$
	\begin{align}
		\xb_0^{k+1}-\xb_0^{k} \to \zerob,~~\xb_i^{k+1}-\xb_i^{k}\to \zerob,\quad \forall~i\in \Vc.
		\label{thm: Lc progress eqn5}
	\end{align}
	Given \eqref{lemma: Lc progress eq 9}, \eqref{thm: Lc progress eqn5} infers
	\begin{align}
		\lambdab_i^{k+1}-\lambdab_i^{k}\to \zerob,\quad \forall i\in \Vc.  \label{thm: Lc progress eqn6}
	\end{align}
	
	We use \eqref{thm: Lc progress eqn5} and \eqref{thm: Lc progress eqn6} to show that every limit point of $(\{\xb_i^k\}_{i=1}^N,\xb_0^k, \{\lambdab_i^k\}_{i=1}^N)$ is a KKT point of problem \eqref{eqn: consensus problem}.
	Firstly, by applying \eqref{thm: Lc progress eqn6}
	to \eqref{eqn: async cadmm s1 lambda equi} and by  \eqref{thm: Lc progress eqn5}, one obtains
	$\xb_0^{k+1}-\xb_i^{k+1} \to \zerob~\forall i\in \Ac_k$. For $i \in \Ac_k^c$, note that $i\in \Ac_{\widetilde k_i}$ (see the definition of $\widetilde k_i$ above \eqref{lemma: Lc progress eq 8}) and thus, by \eqref{eqn: async cadmm s1 lambda equi},
	$$
	\lambdab^{\widetilde k_i+1}_i =\lambdab^{\widetilde k_i}_i + \rho (\xb_i^{\widetilde k_i +1}-\xb_0^{\overline{(\widetilde k_i)}_i +1}),
	$$
	where $\overline{(\widetilde k_i)}_i$ denotes the last iteration number before iteration $\widetilde k_i$ for which worker $i$ is arrived.
	Moreover, since $\xb_i^{\widetilde k_i+1}=\xb_i^{\widetilde k_i+2}=\cdots=\xb_i^{k}=\xb_i^{k+1}$ $\forall i \in \Ac_k^c$, and by \eqref{eqn: async cadmm s1 lambda equi}, \eqref{thm: Lc progress eqn5} and \eqref{thm: Lc progress eqn6}, we have  $\forall i \in \Ac_k^c$,
	\begin{align}
		&\|\xb_0^{k+1}-\xb_i^{k+1}\| = \|\xb_0^{k+1}-\xb_i^{\widetilde k_i+1}\| \notag \\
		&~~= \|\xb_0^{k+1}-\xb_0^{\overline{(\widetilde k_i)}_i +1} + \xb_0^{\overline{(\widetilde k_i)}_i +1} - \xb_i^{\widetilde k_i+1}\|
		\notag \\
		&~~\leq  \|\xb_0^{k+1}-\xb_0^{\overline{(\widetilde k_i)}_i +1}\| + \frac{1}{\rho}\|\lambdab^{\widetilde k_i+1}_i - \lambdab^{\widetilde k_i}_i\|\notag \\
		&~~\to \zerob.
	\end{align}
	So we conclude
	\begin{align}
		\xb_0^{k+1}-\xb_i^{k+1}\to \zerob~\forall i\in \Vc.  \label{thm: Lc progress eqn7}
	\end{align}
	Secondly,
	the optimality condition of \eqref{eqn: async cadmm s1 x0 equi} gives
	\begin{align}
		&{\ssb_0^{k+1}}- \sum_{i=1}^N \lambdab_i^{k+1}
		-\rho \sum_{i=1}^N (\xb_i^{k+1} -\xb_0^{k+1}) \notag \\
		&~~~~~~~~~~~~+
		\gamma (\xb_0^{k+1}-\xb_0^{k}) =\zerob,\label{thm: Lc progress eqn7.5}
	\end{align}
	{for some $\ssb_0^{k+1} \in \partial h(\xb_0^{k+1})$.
		By applying \eqref{thm: Lc progress eqn7} and \eqref{thm: Lc progress eqn5} to \eqref{thm: Lc progress eqn7.5},} we obtain that
	\begin{align}\label{thm: Lc progress eqn8}
		{\ssb_0^{k+1} } - \sum_{i=1}^N \lambdab_i^{k+1} \to \zerob.
	\end{align}
	Equations \eqref{lemma: Lc progress eq 8}, \eqref{thm: Lc progress eqn7} and \eqref{thm: Lc progress eqn8} imply that
	$(\{\xb_i^k\}_{i=1}^N,\xb_0^k, \{\lambdab_i^k\}_{i=1}^N)$ asymptotically satisfy the KKT conditions in \eqref{eqn: KKT of prob}.
	
	{Lastly, let us show that $(\{\xb_i^k\}_{i=1}^N,\xb_0^k, \{\lambdab_i^k\}_{i=1}^N)$ is bounded and has limit points.
		Since dom$(h)$	is compact and $\xb_0^k \in$ dom$(h)$, $\xb_0^k$ is a bounded {sequence} and thus has limit points.
		From \eqref{thm: Lc progress eqn7},  $\xb_i^k$, $i\in \Vc$, are bounded and have limit points. Moreover, by  \eqref{lemma: Lc progress eq 8}, $\lambdab_i^k$, $i\in \Vc$,  are bounded and have limit points as well.  In summary, $(\{\xb_i^k\}_{i=1}^N,\xb_0^k, \{\lambdab_i^k\}_{i=1}^N)$ converges to the set of KKT points of problem \eqref{eqn: consensus problem} .
	}
	\hfill $\blacksquare$
	
	
	{\bf Proof of Corollary \ref{corollary: conv of scheme 1 convex}:} The proof exactly follows that of Theorem \ref{thm: conv of scheme 1}.
	The only difference is that the coefficient of the term $\frac{(1-\rho)+L}{2}\!\sum_{i\in \Ac_k}\!\|\xb_i^{k+1}\! -\!\xb_i^{k}\|^2$ in \eqref{lemma: Lc progress eq 7} reduces from $\frac{(1-\rho)+L}{2}$ to $\frac{(1-\rho)}{2}$; see the footnote in Appendix \ref{appx: proof of lemma: Lc progress}. \hfill $\blacksquare$
	
	\section{Comparison with an Alternative Scheme}\label{sec: alternative scheme}
	
	In Algorithm \ref{table: async cadmm s1 master}, the workers compute $(\xb_i,\lambdab_i),$ $i\in \Vc$, and the master is in charge of computing $\xb_0$. While such distributed implementation is intuitive and natural, one may wonder whether there exist other valid implementations, and if so, how they compare with Algorithm \ref{table: async cadmm s1 master}. To shed some light on this question, we consider in this section an alternative scheme in Algorithm \ref{table: async cadmm s2 master}.

	\begin{algorithm}[h!] \label{table: async cadmm s2}
		\caption{An Alternative Implementation of Asynchronous Distributed ADMM.}
		\begin{algorithmic}[1]\label{table: async cadmm s2 master}
			\STATE {\bf \underline{Algorithm of the Master:}}
			\STATE {\bf Given} initial variable
			$\xb^{0}$ and broadcast it to the workers. Set $k=0$ and $d_1=\cdots=d_N=0;$
			\REPEAT
			\STATE  {\bf wait} {until receiving $\{\hat\xb_i,\hat\lambdab_i\}_{i\in \Ac_k}$ from workers $i\in \Ac_k$ such that  $|\Ac_k|\geq A$ and
				$d_i <\tau-1$ $\forall i \in \Ac_k^c$.}
			\STATE {\bf update}
			\begin{align}&\xb^{k+1}_i =\bigg\{\begin{array}{ll}
					\hat\xb_i  & \forall i\in \Ac_k \\
					\xb^{k}_i & \forall i\in \Ac_k^c
				\end{array},\label{eqn: async cadmm s2 xii} \\
				&d_i =\bigg\{\begin{array}{ll}
					0  & \forall i\in \Ac_k \\
					d_i+1 & \forall i\in \Ac_k^c
				\end{array}, \notag \\
				&\xb^{k+1}_0 \! =\!\arg\min_{\xb_0 \in \mathbb{R}^n} \textstyle  \bigg\{h(\xb_0)-  \xb_0^T\sum_{i=1}^N \lambdab_i^{k} \notag \\
				 &~\textstyle+\frac{\rho}{2}\sum_{i=1}^N\|\xb_i^{k+1}-\xb_0\|^2+\frac{\gamma}{2}\|\xb_0-\xb_0^k\|^2\bigg\},\label{eqn: async cadmm s2 x0}
				\\
				&\lambdab_i^{{k}+1}= \lambdab^{{k}}_i + \rho (\xb_i^{{k_i}+1}-\xb_0^{k+1})~\forall i\in \Vc.\label{eqn: async cadmm s2 lambda}
			\end{align}
			\STATE {\bf broadcast} $\xb^{k+1}_0$ and $\{\lambdab_i^{k+1}\}_{i\in \Ac_k}$ to the workers in $\Ac_k$.
			\STATE {\bf set} $k\leftarrow k+1.$
			\UNTIL {a predefined stopping criterion is satisfied.}%
		\end{algorithmic}
		
		\begin{algorithmic}[1]\label{table: async cadmm s2 worker}
			
			\STATE {\bf \underline{Algorithm of the $i$th Worker:}}
			\STATE {\bf Given} initial $\lambdab^{0}$ and set ${k_i}=0.$
			\REPEAT
			\STATE  {\bf wait} until receiving $(\hat\xb_0,\hat \lambdab_i)$ from the master node.
			\STATE {\bf update}
			\begin{align}
				\xb_i^{{k_i}+1} &=\arg\min_{\xb_i \in \mathbb{R}^n} \textstyle f_i(\xb_i)+\xb_i^T\hat \lambdab_i+\frac{\rho}{2}\|\xb_i -\hat\xb_0\|^2,
				\label{eqn: async cadmm s2 xi}
			\end{align}
			\STATE {\bf send} $\xb_i^{{k_i}+1}$ to the master node.
			\STATE {\bf set} ${k_i}\leftarrow {k_i}+1.$
			\UNTIL {a predefined stopping criterion is satisfied.}%
		\end{algorithmic}
	\end{algorithm}

	Algorithm \ref{table: async cadmm s2 master} differs from Algorithm \ref{table: async cadmm s1 master} in that the master handles not only the update of $\xb_0$ but also that of $\{\lambdab_i\}_{i\in \Vc}$; so the workers only updates $\{\xb_i\}$.
	In essence, in a synchronous network, Algorithm \ref{table: async cadmm s1 master} and Algorithm \ref{table: async cadmm s2 master} are equivalent up to a change of update order\footnote{Algorithm \ref{table: async cadmm s1 master} under the synchronous protocol is the same as Algorithm \ref{table: sync cadmm} with the order of \eqref{eqn: sync cadmm s2} and \eqref{eqn: sync cadmm s1} interchanged.} and have the same convergence conditions. However, intriguingly, in an asynchronous network, the two algorithms may require distinct convergence conditions and behave very differently in practice. To analyze the convergence of Algorithm \ref{table: async cadmm s2 master}, we make the following assumption.
	\begin{Assumption} \label{assumption obj s2}
		Each function $f_i$ is strongly convex with modulus $\sigma^2>0$ and the function $h$ is convex.
	\end{Assumption}
	Under the strong convexity assumption, we are able to show the following convergence result for Algorithm \ref{table: async cadmm s2 master}.
	\begin{Theorem} \label{thm: conv of scheme 2}Suppose that Assumption \ref{assumption bounded delay} and Assumption \ref{assumption obj s2} hold true. Moreover, let $\gamma=0$ and
		\begin{align}\label{eqn: cond of conv scheme2}
			0< \rho \leq
			\frac{\sigma^2}{(5\tau-3)\max\{2\tau,3(\tau-1)\}},
		\end{align} and define $\bar \xb_i^k=\frac{1}{k}\sum_{\ell=1}^k\xb_i^k$ $\forall i=0,1,\ldots,N$, where $(\{\xb_i^k\}_{i=1}^N,\xb_0^k)$ are generated by \eqref{eqn: async cadmm s2 xii} and \eqref{eqn: async cadmm s2 x0}.
		Then, it holds that
		\begin{align}\label{thm: conv of scheme 2 eqn 0}
			\bigg|\bigg[\sum_{i=1}^N f_i(\bar \xb_i^{k})+h(\bar \xb_0^k)\bigg]-F^\star\bigg| &+ \sum_{i=1}^N \|\bar \xb_i^{k}-\bar \xb_0^{k}\| \notag \\
			&\leq \frac{(2+\delta_\lambda)C}{k}
		\end{align} for all $k$, where $C<\infty$ is a finite constant and
		$\delta_\lambda\triangleq \max\{\|\lambdab_1^\star\|,\ldots,\|\lambdab_N^\star\|\}$,
		in which $\{\lambdab_i^\star\}$ denote the optimal dual variables of \eqref{eqn: consensus problem}.%
		%
	\end{Theorem}

	The proof is presented in Appendix \ref{appx: proof of thm conv of sheme 2}.
	Theorem \ref{thm: conv of scheme 2} somehow implies that Algorithm \ref{table: async cadmm s2 master} may require stronger convergence conditions than Algorithm \ref{table: async cadmm s1 master} in the asynchronous network, as $f_i$'s are assumed to be strongly convex. Besides, different from Theorem \ref{thm: conv of scheme 1} where $\rho$ is advised to be large for Algorithm \ref{table: async cadmm s1 master}, Theorem \ref{thm: conv of scheme 2} indicates that $\rho$ needs to be small for Algorithm \ref{table: async cadmm s2 master}. Since $\rho$ is the step size of the dual gradient ascent in \eqref{eqn: async cadmm s2 lambda}, \eqref{eqn: cond of conv scheme2} implies that the master should move $\lambdab_i$'s slowly when $\tau$ is large. {Such insight is reminiscent of  the recent convergence results for multi-block ADMM in \cite{HongLuo2013}.}
	
	
	Interestingly and surprisingly, our numerical results to be presented shortly suggest that the strongly convex $f_i$'s and a small $\rho$ are necessary for the convergence of Algorithm \ref{table: async cadmm s2 master}.
	
	\section{Simulation Results}\label{sec: simu}
	The main purpose of this section is to examine the convergence behavior of the AD-ADMM with respect to the master's iteration number $k$. So, the simulation results to be presented are obtained by implementing Algorithm \ref{table: async cadmm global view} on a desktop computer. First, we present the simulation results of the AD-ADMM for solving the non-convex sparse PCA problem. Second, we consider the LASSO problem and compare Algorithm \ref{table: async cadmm s2 master} with Algorithm \ref{table: async cadmm s1 master}.
	\subsection{Example 1: Sparse PCA}
\begin{figure}[t]\centering
		\resizebox{0.45\textwidth}{!}{
			\includegraphics{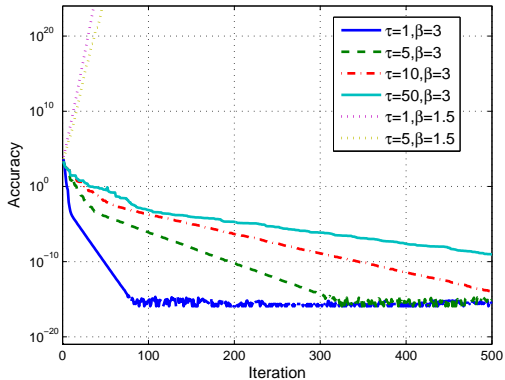}}
		\caption{Convergence curves of the AD-ADMM (Algorithm \ref{table: async cadmm s1 master}) for solving the sparse PCA problem \eqref{spca}; $N=32$, $\theta=0.1$, $\rho=\beta\max_{j=1,\ldots,N} \lambda_{\max}(\Bb_j^T\Bb_j)$ and $\gamma=0.$}
		\vspace{-0.3cm}\label{fig: space}
	\end{figure}

	{Theorem \ref{thm: conv of scheme 1} has} shown that
	the AD-ADMM can converge for non-convex problems. To verify this point, let us consider the following sparse PCA problem \cite{Richtarikspca12}
	\begin{align}\label{spca}
		\!\!\!\!\!\min_{\substack{\wb\in \mathbb{R}^n}}&~-\sum_{j=1}^{N} \wb^T\Bb_j^T\Bb_j\wb \!+\!\theta \|\wb\|_1,
	\end{align}
	where $\Bb_j\in \mathbb{R}^{m\times n}$, $\forall j=1,\ldots,N,$ and $\theta>0$ is a regularization parameter.
	The sparse PCA problem above is not a convex problem. We display in Figure \ref{fig: space} the convergence performance of the AD-ADMM for solving \eqref{spca}. In the simulations, each matrix $\Bb_j \in \mathbb{R}^n$ is a $1000 \times 500$ sparse random matrix with approximately $5000$ non-zero entries; $\theta$ is set to $0.1$ and $N=32.$ The penalty parameter $\rho$ is set to $\rho=\beta\max_{j=1,\ldots,N} \lambda_{\max}(\Bb_j^T\Bb_j)$ and $\gamma=0$. To simulate an asynchronous scenario, at each iteration, half of the workers are assumed to have a probability 0.1 to be arrived independently, and half of the workers are assumed to have a probability 0.8 to be ``arrived" independently. {At each iteration, the master proceeds to update the variables as long as there is at least one arrived worker, i.e., $A=1$.}
	The accuracy is defined as
	\begin{align}\label{eqn: accuracy}
		{\rm accuracy}=\frac{|\Lc_\rho(\xb^k,\xb_0^k,\lambdab^k)-\hat F|}{\hat F}
	\end{align}
	where $\hat F$ denotes the optimal objective value for the synchronous case $(\tau=1)$ which is obtained by running the distributed ADMM (with $\beta=3$) for 10000 iterations (it is found in the experiments that the AD-ADMM converges to the same KKT point for different values of $\tau$).
	One can observe from Figure \ref{fig: space} that the AD-ADMM (with $\beta=3$) indeed converges properly even though \eqref{spca} is a non-convex problem. 
	
	Interestingly, we note that for the example considered here, the AD-ADMM with $\gamma=0$ works well for different values of $\tau$, even though Theorem \ref{thm: conv of scheme 1} suggests that $\gamma$ should be a larger value in the worst-case.
	However, we do observe from Figure \ref{fig: space} that if one sets $\beta=1.5$ (i.e., a smaller value of $\rho$), then the AD-ADMM diverges even in the synchronous case ($\tau=1$). This implies that the claim of a large enough $\rho$ is necessary for the non-convex sparse PCA problem.

	\subsection{Example 2: LASSO}
	
	In this example, we compare the convergence performance of Algorithm \ref{table: async cadmm s2 master} with Algorithm \ref{table: async cadmm s1 master}. We consider the following LASSO problem
	\begin{align}\label{eqn: lasso}
		\min_{\substack{\wb\in \mathbb{R}^n}}~ &\sum_{i=1}^N \|\Ab_i\wb - \bb_i\|^2 + \theta \|\wb\|_1,
	\end{align}
	where $\Ab_i\in \mathbb{R}^{m \times n}$, $\bb_i \in \mathbb{R}^{m},$ $i=1,\ldots,N$, and $\theta>0$.
	The elements of $\Ab_i$'s are randomly generated following the Gaussian distribution with zero mean and unit variance, i.e., $\sim \Nc(0,1)$; each $\bb_i$ is generated by $\bb_i=\Ab_i\wb^0 +\nub_i$ where $\wb^0 \in \mathbb{R}^n$ is an $n \times 1$ sparse random vector with approximately $0.05n$ non-zero entries and $\nub_i$ is a noise vector with entries following $\Nc(0,0.01)$. A star network with 16 ($N=16$) workers is considered.
	To simulate an asynchronous scenario, at each iteration, half of the workers are assumed to have a probability 0.1 to be arrived independently, 4 workers are assumed to have a probability 0.3 to be arrived independently, and the remaining 4 workers are assumed to have a probability 0.8 to be arrived independently.

	\begin{figure*}[!t]
		\begin{center}
			{\subfigure[][Algorithm \ref{table: async cadmm s1 master}, $n=100$]{\resizebox{.47\textwidth}{!}
					{\includegraphics{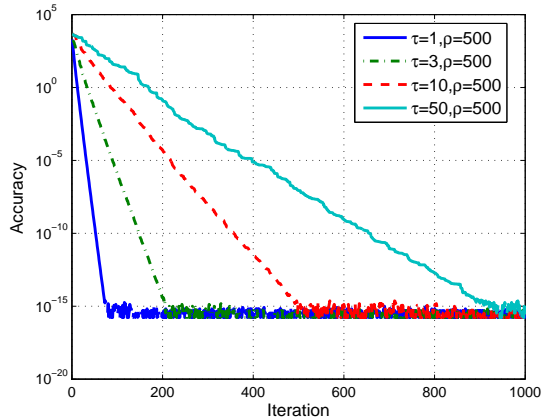}}}
			}
			\hspace{-1pc}
			{\subfigure[][Algorithm \ref{table: async cadmm s2 master}, $n=100$]{\resizebox{.47\textwidth}{!}{\includegraphics{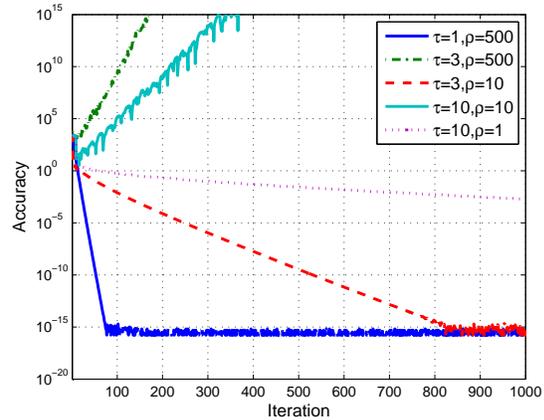}}}
			}
			\hspace{-1pc}
			{\subfigure[][Algorithm \ref{table: async cadmm s1 master}, $n=1000$]{\resizebox{.47\textwidth}{!}{\includegraphics{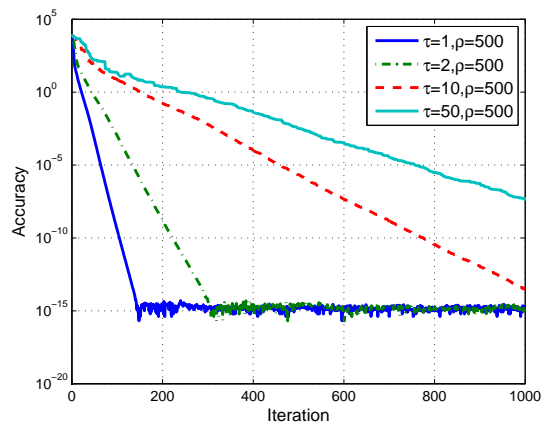}}}
			}
			\hspace{-1pc}
			{\subfigure[][Algorithm \ref{table: async cadmm s2 master}, $n=1000$]{\resizebox{.47\textwidth}{!}
					{\includegraphics{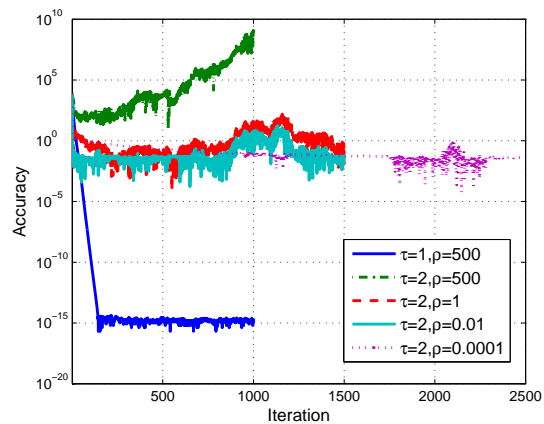}}}
			}
		\end{center}\vspace{-0.0cm}
		\caption{Convergence curves of Algorithm \ref{table: async cadmm s1 master} and Algorithm \ref{table: async cadmm s2 master} for solving the LASSO problem in \eqref{eqn: lasso} with $N=16$, $m=200$ and $\theta=0.1$. The parameter $\gamma$ is set to zero.}
		\vspace{-0.0cm}\label{fig: s1 vs s2 lasso}
	\end{figure*}
	
	Figure \ref{fig: s1 vs s2 lasso}(a) and Figure \ref{fig: s1 vs s2 lasso}(b) respectively display the convergence curves (accuracy versus iteration number) of Algorithm \ref{table: async cadmm s1 master} and Algorithm \ref{table: async cadmm s2 master} for solving \eqref{eqn: lasso} with $N=16$, $m=200$, $n=100$ and $\theta=0.1$. The accuracy is defined as
	\begin{align}
		{\rm accuracy}=\frac{|\Lc_\rho(\xb^k,\xb_0^k,\lambdab^k)-F^\star|}{F^\star}
	\end{align}
	where $F^\star$ denotes the optimal objective value of problem \eqref{eqn: lasso}.
	One can see from Figure \ref{fig: s1 vs s2 lasso}(a) that Algorithm \ref{table: async cadmm s1 master} (with $\rho=500$, $\gamma=0$) converges well for various values of delay $\tau$. From Figure \ref{fig: s1 vs s2 lasso}(b), one can observe that, under the synchronous setting (i.e., $\tau=1$), Algorithm \ref{table: async cadmm s2 master} (with $\rho=500$) exhibits a similar behavior as Algorithm \ref{table: async cadmm s1 master} in Figure \ref{fig: s1 vs s2 lasso}(a). However, under the asynchronous setting of $\tau=3$, Algorithm \ref{table: async cadmm s2 master} (with $\rho=500$) diverges as shown in Figure \ref{fig: s1 vs s2 lasso}(b); Algorithm \ref{table: async cadmm s2 master} can become convergent if one decrease $\rho$ to $10$. Analogously, for $\tau=10$, one has to further reduce $\rho$ to $1$ in order to have Algorithm \ref{table: async cadmm s2 master} convergent. However, the convergence speed of Algorithm \ref{table: async cadmm s2 master} with $\rho=1$ is much slower when comparing to Algorithm \ref{table: async cadmm s1 master} in Figure \ref{fig: s1 vs s2 lasso}(a).
	
	Figure \ref{fig: s1 vs s2 lasso}(c) and Figure \ref{fig: s1 vs s2 lasso}(d) show the comparison results of Algorithm \ref{table: async cadmm s1 master} and Algorithm \ref{table: async cadmm s2 master} for solving \eqref{eqn: lasso} with $n$ increased to $1000$. Note that, given $m=200$ and $n=1000$, the cost functions $f_i(\wb_i)\triangleq \|\Ab_i\wb_i - \bb_i\|^2$ in \eqref{eqn: lasso} are no longer strongly convex. One can observe from Figure \ref{fig: s1 vs s2 lasso}(c) that Algorithm \ref{table: async cadmm s1 master} (with $\rho=500$, $\gamma=0$) still converges properly for various values of $\tau$. However, as one can see from Figure \ref{fig: s1 vs s2 lasso}(d),
	Algorithm \ref{table: async cadmm s2 master} always diverges for various values of $\rho$ even when the delay $\tau$ is as small as two. As a result, the strong convexity assumed in Theorem \ref{thm: conv of scheme 2} may also be necessary in practice. We conclude from these simulation results that Algorithm \ref{table: async cadmm s1 master} significantly outperforms Algorithm \ref{table: async cadmm s2 master} in the asynchronous network, even though the two have the same convergence behaviors in the synchronous network.

	\vspace{-0.2cm}
	\section{Concluding Remarks}\label{sec: conclusions}
	In this paper, we have proposed the AD-ADMM (Algorithm \ref{table: async cadmm s1 master}) aiming at solving large-scale instances of problem \eqref{eqn: original problem} over a star computer network. 
	Under the partially asynchronous model, we have shown (in Theorem \ref{thm: conv of scheme 1}) that the AD-ADMM can deterministically converge to the set of KKT points of problem \eqref{eqn: consensus problem}, even in the absence of convexity of $f_i$'s. We have also compared the AD-ADMM (Algorithm \ref{table: async cadmm s1 master}) with an alternative asynchronous implementation (Algorithm \ref{table: async cadmm s2 master}), and illustrated the interesting fact that a slight modification of the algorithm can significantly change the algorithm convergence conditions/behaviors in the asynchronous setting.
	
	From the presented simulation results, we have observed that the AD-ADMM may exhibit linear convergence for some structured instances of problem \eqref{eqn: original problem}. The conditions under which linear convergence can be achieved are presented in the companion paper \cite{ChangAsyncadmm15_p2}. 
	Numerical results which demonstrate the time efficiency of the proposed AD-ADMM on a high performance computer cluster are also presented in \cite{ChangAsyncadmm15_p2}.

	\vspace{-0.0cm}
	\appendices {\setcounter{equation}{0}
		\renewcommand{\theequation}{A.\arabic{equation}}
		
		%
		%
		\section{Proof of Lemma \ref{lemma: Lc progress}}\label{appx: proof of lemma: Lc progress}
		Notice that
		\begin{align}
			&\Lc_{\rho}(\xb^{k+1},\xb_0^{k+1}, \lambdab^{k+1}) -
			\Lc_{\rho}(\xb^k,\xb_0^k, \lambdab^k) \notag \\
			&= \Lc_{\rho}(\xb^{k+1},\xb_0^{k+1}, \lambdab^{k+1}) -
			\Lc_{\rho}(\xb^{k+1},\xb_0^k, \lambdab^{k+1}) \notag \\
			&~~~+ \Lc_{\rho}(\xb^{k+1},\xb_0^k, \lambdab^{k+1}) -
			\Lc_{\rho}(\xb^{k+1},\xb_0^k, \lambdab^{k}) \notag \\
			&~~~+ \Lc_{\rho}(\xb^{k+1},\xb_0^{k}, \lambdab^{k}) -
			\Lc_{\rho}(\xb^{k},\xb_0^k, \lambdab^{k}). \label{lemma: Lc progress eq 1}
		\end{align}
		{We bound the three pairs of the differences on the right hand side (RHS) of \eqref{lemma: Lc progress eq 1} as follows.}
		{Firstly, since  $-\xb_0^T\sum_{i=1}^N \lambdab_i^{k+1} +\frac{\rho}{2}\sum_{i=1}^N\|\xb_i^{k+1}-\xb_0\|^2+ \frac{\gamma}{2}\|\xb_0-\xb_0^k\|^2$
			in \eqref{eqn: async cadmm s1 x0 equi} is strongly convex with respect to (w.r.t.) $\xb_0$ with modulus $\gamma+N\rho$,
			by \cite[Definition 2.1.2]{BK:Nesterov05}, we have
			\begin{align}
				&\bigg(-  (\xb_0^k)^T\sum_{i=1}^N \lambdab_i^{k+1}+\frac{\rho}{2}\sum_{i=1}^N\|\xb_i^{k+1}-\xb_0^k\|^2\bigg)
				\notag\\
				&-\bigg(-  (\xb_0^{k+1})^T\sum_{i=1}^N \lambdab_i^{k+1} \notag \\
				&~~~~~~~~+\frac{\rho}{2}\sum_{i=1}^N\|\xb_i^{k+1}-\xb_0^{k+1}\|^2+ \frac{\gamma}{2}\|\xb_0^{k+1}-\xb_0^k\|^2\bigg) \notag \\
				&\geq \bigg( -\sum_{i=1}^N \lambdab_i^{k+1} +\rho \sum_{i=1}^N(\xb_0^{k+1} - \xb_i^{k+1})  \notag \\
				&+\gamma (\xb_0^{k+1} - \xb_0^k) \bigg)^T(\xb_0^k -\xb_0^{k+1})+
				\frac{\gamma+N\rho}{2}\|\xb_0^{k+1}-\xb_0^k\|^2.  \label{lemma: Lc progress eq 1.0}
			\end{align}
			By the {optimality} condition of \eqref{eqn: async cadmm s1 x0 equi} and the convexity of $h$, we respectively have
			\begin{align}
				&\bigg( \ssb_0^{k+1}-\sum_{i=1}^N \lambdab_i^{k+1} +\rho \sum_{i=1}^N(\xb_0^{k+1} - \xb_i^{k+1})  \notag \\
				&~~~~~~~~~~~~~~~~~+ \gamma (\xb_0^{k+1} - \xb_0^k) \bigg)^T(\xb_0^k -\xb_0^{k+1})\geq 0,   \label{lemma: Lc progress eq 1.1} \\
				&  h(\xb_0^{k})  \geq  h(\xb_0^{k+1}) +   (\ssb_0^{k+1})^T(\xb_0^{k} -\xb_0^{k+1}). \label{lemma: Lc progress eq 1.2}
			\end{align}
			By subsequently applying \eqref{lemma: Lc progress eq 1.1} and \eqref{lemma: Lc progress eq 1.2} to \eqref{lemma: Lc progress eq 1.0}, we obtain
		}
		\begin{align}
			&\bigg(h(\xb_0^k)-  (\xb_0^k)^T\sum_{i=1}^N \lambdab_i^{k+1}+\frac{\rho}{2}\sum_{i=1}^N\|\xb_i^{k+1}-\xb_0^k\|^2\bigg)
			\notag\\
			&-\bigg(h(\xb_0^{k+1})-  (\xb_0^{k+1})^T\sum_{i=1}^N \lambdab_i^{k+1} \notag \\
			&~~~~~~~~+\frac{\rho}{2}\sum_{i=1}^N\|\xb_i^{k+1}-\xb_0^{k+1}\|^2+ \frac{\gamma}{2}\|\xb_0^{k+1}-\xb_0^k\|^2\bigg) \notag \\
			&\geq \frac{\gamma+N\rho}{2}\|\xb_0^{k+1}-\xb_0^k\|^2,
		\end{align} that is,
		\begin{align}\label{eqn: L-L1}
			&\Lc_{\rho}(\xb^{k+1},\xb_0^{k+1}, \lambdab^{k+1}) -
			\Lc_{\rho}(\xb^{k+1},\xb_0^k, \lambdab^{k+1})
			\notag \\
			&\leq -\frac{2\gamma+N\rho}{2}\|\xb_0^{k+1}-\xb_0^k\|^2.
		\end{align}

		Secondly, it directly follows from \eqref{eqn: Lc} that
		\begin{align}\label{eqn: L-L2}
			&\Lc_{\rho}(\xb^{k+1},\xb_0^k, \lambdab^{k+1}) -
			\Lc_{\rho}(\xb^{k+1},\xb_0^k, \lambdab^{k}) \notag \\
			&=\sum_{i=1}^N (\lambdab_i^{k+1}-\lambdab_i^k)^T(\xb_i^{k+1} -\xb_0^k) \notag \\
			&=\sum_{i\in \Ac_k} (\lambdab_i^{k+1}-\lambdab_i^k)^T(\xb_i^{k+1} -\xb_0^{\bar k_i +1})\notag \\
			&~~~~~~~~~~~~~~~~~
			+\sum_{i\in \Ac_k} (\lambdab_i^{k+1}-\lambdab_i^k)^T(\xb_0^{\bar k_i +1} -\xb_0^k) \notag
			\\
			&=\frac{1}{\rho}\sum_{i\in \Ac_k} \|\lambdab_i^{k+1}-\lambdab_i^k\|^2
			\notag \\
			&~~~~~~~~~~~~~~~~~+\sum_{i\in \Ac_k} (\lambdab_i^{k+1}-\lambdab_i^k)^T(\xb_0^{\bar k_i +1} -\xb_0^k),
		\end{align}
		where the second equality is due to the fact that $\lambdab_i^{k+1}=\lambdab_i^k~\forall i\in \Ac_k^c$ and the last equality is obtained by applying
		\begin{align}\label{lemma: Lc progress eq 3.2}
			\lambdab^{k+1}_i=\lambdab^{k}_i + \rho (\xb_i^{k+1}-\xb_0^{\bar k_i +1})~ \forall i\in \Ac_k
		\end{align}
		as shown in \eqref{eqn: async cadmm s1 lambda equi}.
		
		Thirdly, define ${\Lc_i(\xb_i,\xb_0^k,\lambdab^k)}=f_i(\xb_i)+\xb_i^T\lambdab^{k}_i +\frac{\rho}{2}\|\xb_i -\xb_0^{k}\|^2$ and assume that $\rho\geq L$. Since, by \cite[Lemma 1.2.2]{BK:Nesterov05}, the minimum eigenvalue of the Hessian matrix of $f_i(\xb_i)$ is no smaller than $-L$, ${\Lc_i(\xb_i,\xb_0^k,\lambdab^k)}$ is strongly convex w.r.t. $\xb_i$ and the convexity parameter is given by $\rho -L \geq 0$ \footnote{When $f_i$ is a convex function, the minimum eigenvalue of the Hessian matrix of $f_i(\xb_i)$ is zero. So, the convexity parameter of $\Lc_i(\xb_i,\lambdab^k,\xb_0^k)$ is $\rho$ instead.}. Therefore, one has
		\begin{align}
			&{\Lc_i(\xb_i^k,\xb_0^k,\lambdab^k) \geq \Lc_i(\xb_i^{k+1},\xb_0^k,\lambdab^k)}\notag \\
			&~~~~~~+
			(\nabla f_i(\xb_i^{k+1}) + \lambdab^{k}_i + \rho(\xb_i^{k+1} -\xb_0^{k}) )^T(\xb^k_i-\xb_i^{k+1}) \notag \\
			&~~~~~~+\frac{\rho-L}{2}\|\xb_i^{k+1} -\xb_i^{k}\|^2. \label{lemma: Lc progress eq 4}
		\end{align}
		Also, by the optimality condition of \eqref{eqn: async cadmm s1 xi equi}, one has, $\forall i\in \Ac_k$,
		\begin{align}
			\zerob&=\nabla f_i(\xb_i^{k+1}) + \lambdab^{k}_i + \rho(\xb_i^{k+1} -\xb_0^{\bar k_i+1})
			\label{lemma: Lc progress eq 4.5}
			\end{align}
			\begin{align}
			&=
			(\nabla f_i(\xb_i^{k+1}) + \lambdab^{k}_i + \rho(\xb_i^{k+1} -\xb_0^{k}))  \notag \\
			&~~~~~~~~~~~~~~~~~~+ \rho(\xb_0^{k} -\xb_0^{\bar k_i+1}). \label{lemma: Lc progress eq 5}
		\end{align}
		By substituting \eqref{lemma: Lc progress eq 5} into \eqref{lemma: Lc progress eq 4} and by \eqref{eqn: Lc}, we have
		\begin{align}
			&\Lc_{\rho}(\xb^{k+1},\xb_0^{k}, \lambdab^{k}) -
			\Lc_{\rho}(\xb^{k},\xb_0^k, \lambdab^{k}) \notag \\
			&=\sum_{i=1}^N (\Lc_i(\xb_i^{k+1},\lambdab^k,\xb_0^k)-\Lc_i(\xb_i^k,\lambdab^k,\xb_0^k))\notag \\
			&=\sum_{i\in \Ac_k} (\Lc_i(\xb_i^{k+1},\lambdab^k,\xb_0^k)-\Lc_i(\xb_i^k,\lambdab^k,\xb_0^k))\notag \\
			&\leq -\frac{\rho-L}{2}\sum_{i\in \Ac_k}\|\xb_i^{k+1} -\xb_i^{k}\|^2 \notag \\
			&~~~~~~~~+ \rho\sum_{i\in \Ac_k}(\xb_0^{\bar k_i+1}-\xb_0^{k}) ^T(\xb_i^{k+1}-\xb^k_i),
			\label{lemma: Lc progress eq 6}
		\end{align}
		where the second equality is due to $\xb_i^{k+1}=\xb_i^k~\forall i\in \Ac_k^c$ from \eqref{eqn: async cadmm s1 xi equi}.
		
		After substituting \eqref{eqn: L-L1}, \eqref{eqn: L-L2} and \eqref{lemma: Lc progress eq 6} into \eqref{lemma: Lc progress eq 1}, we obtain
		\begin{align}
			&\Lc_{\rho}(\xb^{k+1},\xb_0^{k+1}, \lambdab^{k+1}) -
			\Lc_{\rho}(\xb^k,\xb_0^k, \lambdab^k) \notag \\
			&\leq -\frac{2\gamma+N\rho}{2}\|\xb_0^{k+1}-\xb_0^k\|^2 +\frac{1}{\rho}\sum_{i\in \Ac_k} \|\lambdab_i^{k+1}-\lambdab_i^k\|^2 \notag \\
			&~~~-\!\frac{\rho-L}{2}\!\sum_{i\in \Ac_k}\!\|\xb_i^{k+1}\!-\!\xb_i^{k}\|^2 \notag \\
			&
			~~~+\!\sum_{i\in \Ac_k} \!(\lambdab_i^{k+1}-\lambdab_i^k)^T(\xb_0^{\bar k_i +1} -\xb_0^k) \notag \\
			&~~~ +\! \rho\!\sum_{i\in \Ac_k}(\xb_0^{\bar k_i+1}\!-\!\xb_0^{k}) ^T(\xb_i^{k+1}-\xb^k_i).\label{lemma: Lc progress eq 7.1}
		\end{align}
		Recall the Young's inequality, i.e.,
		\begin{align}
			\ab^T\bb \leq \frac{1}{2\delta}\|\ab\|^2 + \frac{\delta}{2}\|\bb\|^2,
		\end{align}
		for any $\ab$, $\bb$ and $\delta>0$, and apply it to the {fourth} and fifth terms in the RHS of \eqref{lemma: Lc progress eq 7.1} with $\delta=1$ and $ \delta=1/\rho$ for some $\epsilon>0$, respectively. Then \eqref{lemma: Lc progress eq 7} is obtained. \hfill $\blacksquare$

		\section{Proof of Lemma \ref{lemma: asyn error bound}}\label{appx: proof of lemma: asyn error bound}
		It is easy to show that
		\begin{align}\label{lemma: asyn error bound eq1}
			&\sum_{j=0}^{k}\sum_{i\in \Ac_j} \|\xb_0^{j}-\xb_0^{\bar j_i+1}\|^2
			=\sum_{j=0}^{k}\sum_{i\in \Ac_j} \|\sum_{\ell=\bar j_i+1}^{j-1}(\xb_0^{\ell}-\xb_0^{\ell+1})\|^2
			\notag \\
			&\leq \sum_{j=0}^{k}\sum_{i\in \Ac_j}(j-\bar j_i -1)\sum_{\ell=\bar j_i+1}^{j-1} \|\xb_0^{\ell}-\xb_0^{\ell+1}\|^2
			\notag \\
			&\leq \sum_{j=0}^{k}\sum_{i\in \Ac_j}(\tau-1)\sum_{\ell=j-\tau+1}^{j-1} \|\xb_0^{\ell}-\xb_0^{\ell+1}\|^2 \notag \\
			&\leq S(\tau-1)\sum_{j=0}^{k}\sum_{\ell=j-\tau+1}^{j-1} \|\xb_0^{\ell}-\xb_0^{\ell+1}\|^2
		\end{align}
		where, in the second inequality, we have applied the fact of $j-\tau\leq \bar j_i <j$ from \eqref{eqn: cond on bar ki}; in the last inequality, we have applied the assumption of $|\Ac_k|<S$ for all $k$.
		Notice that, in the summation $\sum_{j=0}^{k}\sum_{\ell=j-\tau+1}^{j-1} \|\xb_0^{\ell}-\xb_0^{\ell+1}\|^2$, each $\|\xb_0^{j}-\xb_0^{j+1}\|^2$, where $j=0,\ldots,k-1$, appears no more than $\tau-1$ times. Thus, one can upper bound
		\begin{align}
			\sum_{j=0}^{k}\sum_{\ell=j-\tau+1}^{j-1} \|\xb_0^{\ell}-\xb_0^{\ell+1}\|^2
			&\leq (\tau-1)\sum_{j=0}^{k-1}\|\xb_0^{j+1}-\xb_0^{j}\|^2, \label{thm: Lc progress eqn3}
		\end{align}
		which, combined with \eqref{lemma: asyn error bound eq1}, yields \eqref{lemma: asyn error bound eq0}.\hfill $\blacksquare$
		
		\section{Proof of Lemma \ref{lemma: Lc lower bounded}}\label{appx: proof of lemma: Lc lower bounded}
		The proof is similar to \cite[Lemma 2.3]{Hong15noncvxadmm}. We present the proof here for completeness.
		By recalling equation \eqref{lemma: Lc progress eq 8} and applying it to \eqref{eqn: Lc}, one obtains
		\begin{align}\label{eqn: proof of claim 1 2}
			&\Lc_{\rho}(\xb^{k+1},\xb_0^{k+1}, \lambdab^{k+1})
			= h(\xb_0^{k+1})+\sum_{i=1}^N f_i(\xb_i^{k+1}) \notag \\
			&-\sum_{i=1}^N (\nabla f_i(\xb_i^{k+1}))^T(\xb_i^{k+1} -\xb_0^{k+1}) +\frac{\rho}{2}\sum_{i=1}^N\|\xb_i^{k+1} -\xb_0^{k+1}\|^2.
		\end{align}
		As $\nabla f_i$ is {Lipschitz} continuous under Assumption \ref{assumption obj s1}, the descent lemma \cite[Proposition A.24]{BK:Bertsekas2003_NP} holds
		\begin{align}\label{eqn: proof of claim 1 3}
			f_i(\xb_0^{k+1})&\leq f_i(\xb_i^{k+1}) + (\nabla f_i(\xb_i^{k+1}))^T(\xb_0^{k+1}-\xb_i^{k+1})\notag \\&+\frac{L}{2}\|\xb_i^{k+1}-\xb_0^{k+1}\|^2~\forall~i=1,\ldots,N.
		\end{align}
		By combining \eqref{eqn: proof of claim 1 2} and \eqref{eqn: proof of claim 1 3}, one can lower bound $\Lc_{\rho}(\xb^{k+1},\xb_0^{k+1}, \lambdab^{k+1})$ as
		\begin{align}\label{eqn: proof of claim 1 4}
			&\Lc_{\rho}(\xb^{k+1},\xb_0^{k+1}, \lambdab^{k+1})\geq  h(\xb_0^{k+1})+\sum_{i=1}^N f_i(\xb_0^{k+1}) \notag \\
			&~~~~~~~+\frac{\rho-L}{2}\sum_{i=1}^N\|\xb_i^{k+1} -\xb_0^{k+1}\|^2,
		\end{align}
		which implies \eqref{eqn: Lc lower bounded} given $\rho\geq L$ and under Assumption \ref{assumption obj s1}. \hfill $\blacksquare$

\section{Proof of Theorem \ref{thm: conv of scheme 2}}\label{appx: proof of thm conv of sheme 2}
For ease of analysis, we equivalently write Algorithm \ref{table: async cadmm s2 master} as follows:
For iteration $k=0,1,\ldots,$
\begin{align}\xb^{k+1}_i &=\bigg\{\begin{array}{ll}
		\arg~{\displaystyle \min_{\xb_i}}~ \textstyle f_i(\xb_i)+\xb_i^T\lambdab^{\bar k_i +1}_i +\frac{\rho}{2}\|\xb_i -\xb_0^{\bar k_i +1}\|^2,
		& \forall i\in \Ac_k \\
		\xb^{k}_i & \forall i\in \Ac_k^c
	\end{array},\label{eqn: async cadmm s2 xi equi} \\
	\xb^{k+1}_0 & =\arg~\min_{\xb_0} ~ \textstyle  h(\xb_0)-  \xb_0^T\sum_{i=1}^N \lambdab_i^{k}+\frac{\rho}{2}\sum_{i=1}^N\|\xb_i^{k+1}-\xb_0\|^2,
	\label{eqn: async cadmm s2 x0 equi}\\
	\lambdab_i^{k+1}&= \lambdab^{k}_i + \rho (\xb_i^{k+1}-\xb^{k+1}_0 )~\forall i\in \Vc.
	\label{eqn: async cadmm s2 lambda equi}
\end{align}
Here, $\bar k_i$ is the last iteration number for which the master node receives message from worker $i\in \Ac_k$ before iteration $k$. For $i\in \Ac_k^c$, let us denote $\widetilde k_i$ ($k-\tau<\widetilde k_i<k$) as the last iteration number for which the master node receives message from worker $i$ before iteration $k$, and further denote $\widehat k_i$ ($\widetilde k_i-\tau \leq \widehat k_i <\widetilde k_i$) as the last iteration number for which the master node receives message from worker $i$ before iteration $\widetilde k_i$. Then, by \eqref{eqn: async cadmm s2 xi equi}, it must be
\begin{align}
	\xb^{\widetilde k_i+1}_i &=\arg~{\displaystyle \min_{\xb_i}}~ \textstyle f_i(\xb_i)+\xb_i^T\lambdab^{\widehat k_i +1}_i +\frac{\rho}{2}\|\xb_i -\xb_0^{\widehat k_i+1}\|^2~~
	\forall i\in \Ac_k^c, \label{eqn: async cadmm s2 xi skc equi}\\
	\xb^{k+1}_i&=\xb^{\widetilde k_i+1}_i, \label{eqn: async cadmm s2 xi skc equi2}
\end{align}
where the second equation is due to $\xb^{\widetilde k_i+1}_i=\xb^{\widetilde k_i+2}_i=\cdots=\xb^{k}_i=\xb^{k+1}_i~\forall i\in \Ac_k^c$.

Let us consider the following update steps
\begin{align}\xb^{k+1}_i &=\bigg\{\begin{array}{ll}
		\arg~{\displaystyle \min_{\xb_i}}~ \textstyle  \alpha f_i(\xb_i)+\xb_i^T\widetilde \lambdab^{\bar k_i +1}_i +\frac{\beta}{2}\|\xb_i -\xb_0^{\bar k_i +1}\|^2,
		& \forall i\in \Ac_k \\
		\arg~{\displaystyle \min_{\xb_i}}~ \textstyle \alpha f_i(\xb_i)+\xb_i^T\widetilde \lambdab^{\widehat k_i +1}_i +\frac{\beta}{2}\|\xb_i -\xb_0^{\widehat k_i+1}\|^2 & \forall i\in \Ac_k^c
	\end{array},\label{eqn: async cadmm s2 xi equi2} \\
	\xb^{k+1}_0 & =\arg~\min_{\xb_0} ~ \textstyle  \alpha h(\xb_0)-  \xb_0^T\sum_{i=1}^N \widetilde \lambdab_i^{k}+\frac{\beta}{2}\sum_{i=1}^N\|\xb_i^{k+1}-\xb_0\|^2,
	\label{eqn: async cadmm s2 x0 equi2}\\
	\widetilde \lambdab_i^{k+1}&= \widetilde \lambdab^{k}_i + \beta (\xb_i^{k+1}-\xb^{k+1}_0 )~\forall i\in \Vc,
	\label{eqn: async cadmm s2 lambda equi2}
\end{align} where $\alpha,\beta>0$. One can verify that \eqref{eqn: async cadmm s2 xi equi2}-\eqref{eqn: async cadmm s2 lambda equi2} are equivalent to \eqref{eqn: async cadmm s2 xi equi}-\eqref{eqn: async cadmm s2 lambda equi} and \eqref{eqn: async cadmm s2 xi skc equi}-\eqref{eqn: async cadmm s2 xi skc equi2} if one considers the change of variables $\lambdab_i=\widetilde \lambdab_i/\alpha$ and $\rho=\beta/\alpha$.

We first consider the optimality condition of \eqref{eqn: async cadmm s2 xi equi2} for $i\in \Ac_k$:
\begin{align}
	0&\geq  \alpha\partial  f_i(\xb_i^{k+1})^T(\xb_i^{k+1}-\xb_i^\star) + (\widetilde \lambdab^{\bar k_i+1}_i + \beta(\xb_i^{k+1} -\xb_0^{\bar k_i+1}))^T(\xb_i^{k+1}-\xb_i^\star) \notag \\
	&=\alpha\partial  f_i(\xb_i^{k+1})^T(\xb_i^{k+1}-\xb_i^\star) + (\widetilde \lambdab^{k+1}_i)^T(\xb_i^{k+1}-\xb_i^\star)
	\notag \\
	&~~~~~+(\widetilde \lambdab^{\bar k_i+1}_i-\widetilde \lambdab^{k}_i)^T(\xb_i^{k+1}-\xb_i^\star)+\beta(\xb_0^{k+1}-\xb_0^{\bar k_i+1})^T(\xb_i^{k+1}-\xb_i^\star),\label{thm: conv of scheme 2 eqn 1}
\end{align}
where we have applied \eqref{eqn: async cadmm s2 lambda equi2} to obtain the equality. Since, under Assumption 3, $f_i$ is strongly convex, one has
\begin{align}
	\alpha f_i(\xb_i^\star) \geq \alpha f_i(\xb_i^{k+1}) +
	\alpha\partial  f_i(\xb_i^{k+1})^T(\xb_i^\star-\xb_i^{k+1}) +\frac{\alpha\sigma^2}{2}\|\xb_i^{k+1}-\xb_i^\star\|^2.
	\label{thm: conv of scheme 2 eqn 2}
\end{align}
Combining \eqref{thm: conv of scheme 2 eqn 1} and \eqref{thm: conv of scheme 2 eqn 2} gives rise to
\begin{align}
	&\alpha f_i(\xb_i^{k+1})-\alpha f_i(\xb_i^\star)+\widetilde \lambdab_i^T(\xb_i^{k+1}-\xb_i^\star)+\frac{\alpha\sigma^2}{2}\|\xb_i^{k+1}-\xb_i^\star\|^2 \notag \\
	& ~~+(\widetilde \lambdab_i^{k+1}-\widetilde \lambdab_i)^T(\xb_i^{k+1}-\xb_i^\star)+(\widetilde \lambdab^{\bar k_i+1}_i-\widetilde \lambdab^{k}_i)^T(\xb_i^{k+1}-\xb_i^\star)\notag \\
	&~~+\beta(\xb_0^{k+1}-\xb_0^{\bar k_i+1})^T(\xb_i^{k+1}-\xb_i^\star)\leq 0~~\forall i\in \Ac_k.
	\label{thm: conv of scheme 2 eqn 3}
\end{align}
On the other hand, consider the optimality condition of \eqref{eqn: async cadmm s2 xi equi2} for $i\in \Ac_k^c$:
\begin{align}
	0&\geq  \alpha\nabla f_i(\xb_i^{k+1})^T(\xb_i^{k+1}-\xb_i^\star) + (\widetilde \lambdab^{\widehat k_i+1}_i + \beta(\xb_i^{k+1} -\xb_0^{\widehat k_i+1}))^T(\xb_i^{k+1}-\xb_i^\star) \notag \\
	&=\alpha\nabla f_i(\xb_i^{k+1})^T(\xb_i^{k+1}-\xb_i^\star) \notag \\
	&~~~~+ (\widetilde \lambdab^{\widehat k_i+1}_i + \widetilde \lambdab_i^{\widetilde k_i+1}- \widetilde \lambdab^{\widetilde k_i}_i - \beta (\xb_i^{\widetilde k_i+1}-\xb^{\widetilde k_i+1}_0 ) + \beta(\xb_i^{\widetilde k_i+1} -\xb_0^{\widehat k_i+1}))^T(\xb_i^{k+1}-\xb_i^\star) \notag \\
	&=\alpha\nabla f_i(\xb_i^{k+1})^T(\xb_i^{k+1}-\xb_i^\star) + (\widetilde \lambdab^{\widetilde k_i+1}_i)^T(\xb_i^{k+1}-\xb_i^\star)
	\notag \\
	&~~~~~+(\widetilde \lambdab^{\widehat k_i+1}_i-\widetilde \lambdab^{\widetilde k_i})^T(\xb_i^{k+1}-\xb_i^\star)+\beta(\xb_0^{\widetilde k_i+1}-\xb_0^{\widehat k_i+1})^T(\xb_i^{k+1}-\xb_i^\star),
	\label{thm: conv of scheme 2 eqn 4}
\end{align}
where \eqref{eqn: async cadmm s2 lambda equi2} with $k=\widetilde k_i$ and \eqref{eqn: async cadmm s2 xi skc equi2} are used to obtain the first equality.
By combining \eqref{thm: conv of scheme 2 eqn 2} with \eqref{thm: conv of scheme 2 eqn 4}, one obtains
\begin{align}
	&\alpha f_i(\xb_i^{k+1})-\alpha f_i(\xb_i^\star)+\widetilde \lambdab_i^T(\xb_i^{k+1}-\xb_i^\star)+\frac{\alpha\sigma^2}{2}\|\xb_i^{k+1}-\xb_i^\star\|^2 \notag \\
	& ~~+(\widetilde \lambdab_i^{\widetilde k_i+1}-\widetilde \lambdab_i)^T(\xb_i^{k+1}-\xb_i^\star)+(\widetilde \lambdab^{\widehat k_i+1}_i-\widetilde \lambdab^{\widetilde k_i})^T(\xb_i^{k+1}-\xb_i^\star)\notag \\
	&~~+\beta(\xb_0^{\widetilde k_i+1}-\xb_0^{\widehat k_i+1})^T(\xb_i^{k+1}-\xb_i^\star)\leq 0~~\forall i\in \Ac_k^c.
	\label{thm: conv of scheme 2 eqn 5}
\end{align}
By summing \eqref{thm: conv of scheme 2 eqn 3} for all $i\in \Ac_k$ and \eqref{thm: conv of scheme 2 eqn 5} for all
$i\in \Ac_k^c$ and further summing the resultant two terms, we obtain that
\begin{align}
	&\alpha \sum_{i=1}^N f_i(\xb_i^{k+1})-\alpha \sum_{i=1}^N f_i(\xb_i^\star)+\sum_{i=1}^N \widetilde \lambdab_i^T(\xb_i^{k+1}-\xb_i^\star)+\sum_{i=1}^N \frac{\alpha\sigma^2}{2}\|\xb_i^{k+1}-\xb_i^\star\|^2 \notag \\
	& ~~ +\underbrace{\sum_{i\in\Ac_k}(\widetilde \lambdab_i^{k+1}-\widetilde \lambdab_i)^T(\xb_i^{k+1}-\xb_i^\star)
		+\sum_{i\in\Ac_k^c}(\widetilde \lambdab_i^{\widetilde k_i+1}-\widetilde \lambdab_i)^T(\xb_i^{k+1}-\xb_i^\star)}_{\rm (a)}
	\notag\\
	&~~+\sum_{i\in\Ac_k}(\widetilde \lambdab^{\bar k_i+1}_i-\widetilde \lambdab^{k}_i)^T(\xb_i^{k+1}-\xb_i^\star)
	+\sum_{i\in\Ac_k^c}(\widetilde \lambdab^{\widehat k_i+1}_i-\widetilde \lambdab^{\widetilde k_i})^T(\xb_i^{k+1}-\xb_i^\star)\notag \\
	&~~+\underbrace{\sum_{i\in\Ac_k}\beta(\xb_0^{k+1}-\xb_0^{\bar k_i+1})^T(\xb_i^{k+1}-\xb_i^\star)
		+\sum_{i\in\Ac_k^c}\beta(\xb_0^{\widetilde k_i+1}-\xb_0^{\widehat k_i+1})^T(\xb_i^{k+1}-\xb_i^\star)}_{\rm (b)}\leq 0.
	\label{thm: conv of scheme 2 eqn 6}
\end{align}
The term (a) in \eqref{thm: conv of scheme 2 eqn 6}, after adding and subtracting $\sum_{i\in\Ac_k^c}(\widetilde \lambdab_i^{k+1}-\widetilde \lambdab_i)^T(\xb_i^{k+1}-\xb_i^\star)$, can be written as
\begin{align}
	{\rm (a)}=
	\sum_{i=1}^N(\widetilde \lambdab_i^{k+1}-\widetilde \lambdab_i)^T(\xb_i^{k+1}-\xb_i^\star)
	+\sum_{i\in\Ac_k^c}(\widetilde \lambdab_i^{\widetilde k_i+1}-\widetilde \lambdab_i^{k+1})^T(\xb_i^{k+1}-\xb_i^\star).
	\label{thm: conv of scheme 2 eqn 7}
\end{align}
The term (b) in \eqref{thm: conv of scheme 2 eqn 6} can be expressed as
\begin{align}\label{thm: conv of scheme 2 eqn 69}
	{\rm (b)}&=
	\sum_{i\in\Ac_k}\beta(\xb_0^{k+1}-\xb_0^{k}+\xb_0^{k}-\xb_0^{\bar k_i+1})^T(\xb_i^{k+1}-\xb_i^\star)
	+\sum_{i\in\Ac_k^c}\beta(\xb_0^{\widetilde k_i+1}-\xb_0^{\widehat k_i+1})^T(\xb_i^{k+1}-\xb_i^\star)\notag\\
	&=\sum_{i=1}^N\beta(\xb_0^{k+1}-\xb_0^{k})^T(\xb_i^{k+1}-\xb_i^\star)
	+\sum_{i\in\Ac_k^c}\beta(\xb_0^{\widetilde k_i+1}-\xb_0^{\widehat k_i+1}-\xb_0^{k+1}+\xb_0^{k})^T(\xb_i^{k+1}-\xb_i^\star)\notag\\
	&~~~+\sum_{i\in\Ac_k}\beta(\xb_0^{k}-\xb_0^{\bar k_i+1})^T(\xb_i^{k+1}-\xb_i^\star).
\end{align}
Note that, by applying \eqref{eqn: async cadmm s2 lambda equi2} and the fact of $\xb_i^\star=\xb_0^\star~\forall i\in \Vc$, one can write
\begin{align}
	\sum_{i=1}^N\beta(\xb_0^{k+1}-\xb_0^{k})^T(\xb_i^{k+1}-\xb_i^\star)&=\sum_{i=1}^N\beta(\xb_0^{k+1}-\xb_0^{k})^T(\xb_i^{k+1}-\xb_0^{k+1}+\xb_0^{k+1}-
	\xb_i^\star)\notag \\
	&=\sum_{i=1}^N(\xb_0^{k+1}-\xb_0^{k})^T(\widetilde \lambdab_i^{k+1}-\widetilde \lambdab^{k}_i) +N\beta(\xb_0^{k+1}-\xb_0^{k})^T(\xb^{k+1}_0-\xb_0^\star).
\end{align}
So, The term (b) in \eqref{thm: conv of scheme 2 eqn 69} is given by
\begin{align}
	{\rm (b)}&=\sum_{i=1}^N(\xb_0^{k+1}-\xb_0^{k})^T(\widetilde \lambdab_i^{k+1}-\widetilde \lambdab^{k}_i) +N\beta(\xb_0^{k+1}-\xb_0^{k})^T(\xb^{k+1}_0-\xb_0^\star)
	\notag\\
	&~~~+\sum_{i\in\Ac_k^c}\beta(\xb_0^{\widetilde k_i+1}-\xb_0^{\widehat k_i+1}-\xb_0^{k+1}+\xb_0^{k})^T(\xb_i^{k+1}-\xb_i^\star)+\sum_{i\in\Ac_k}\beta(\xb_0^{k}-\xb_0^{\bar k_i+1})^T(\xb_i^{k+1}-\xb_i^\star).
	\label{thm: conv of scheme 2 eqn 8}
\end{align}

It can be shown that
\begin{align}
	\sum_{i=1}^N(\xb_0^{k+1}-\xb_0^{k})^T(\widetilde \lambdab_i^{k+1}-\widetilde \lambdab^{k}_i)\geq 0.\label{thm: conv of scheme 2 eqn 9}
\end{align}
To see this, consider the optimality condition of \eqref{eqn: async cadmm s2 x0 equi2}: $\forall \xb_0\in \mathbb{R}^n$,
\begin{align}
	0&\geq \alpha h(\xb_0^{k+1})- \alpha h(\xb_0)   -\sum_{i=1}^N (\widetilde \lambdab_i^{k}+ \beta (\xb_i^{k+1}-\xb_0^{k+1}))^T(\xb_0^{k+1}-\xb_0) \notag \\
	&=\alpha h(\xb_0^{k+1})- \alpha h(\xb_0)   -\sum_{i=1}^N (\widetilde \lambdab_i^{k+1})^T(\xb_0^{k+1}-\xb_0),
	\label{thm: conv of scheme 2 eqn 10}
\end{align}
where the equality is due to \eqref{eqn: async cadmm s2 lambda equi2}. By letting $\xb_0=\xb_0^k$ in \eqref{thm: conv of scheme 2 eqn 10} and also considering \eqref{thm: conv of scheme 2 eqn 10} for iteration $k$ and $\xb_0=\xb_0^{k+1}$, we have \begin{align}
	0&\geq \alpha h(\xb_0^{k+1})- \alpha h(\xb_0^{k})   -\sum_{i=1}^N (\widetilde \lambdab_i^{k+1})^T(\xb_0^{k+1}-\xb_0^{k}),\notag\\
	0&\geq \alpha h(\xb_0^{k})- \alpha h(\xb_0^{k+1})   -\sum_{i=1}^N (\widetilde \lambdab_i^{k})^T(\xb_0^{k}-\xb_0^{k+1}),
	\label{thm: conv of scheme 2 eqn 11}
\end{align}
respectively. By summing the above two equations, we obtain \eqref{thm: conv of scheme 2 eqn 9}. Moreover, by letting $\xb_0=\xb_i^\star=\xb_0^\star$ in \eqref{thm: conv of scheme 2 eqn 10}, we have
\begin{align}
	\alpha h(\xb_0^{k+1})- \alpha h(\xb_0^\star) -\sum_{i=1}^N \widetilde \lambdab_i^T(\xb_0^{k+1}-\xb_i^\star)
	-\sum_{i=1}^N (\widetilde \lambdab_i^{k+1}-\widetilde \lambdab_i)^T(\xb_0^{k+1}-\xb_i^\star)\leq 0.
	\label{thm: conv of scheme 2 eqn 12}
\end{align}
By summing \eqref{thm: conv of scheme 2 eqn 12} and \eqref{thm: conv of scheme 2 eqn 6} followed by applying \eqref{thm: conv of scheme 2 eqn 7}, \eqref{thm: conv of scheme 2 eqn 8} and \eqref{thm: conv of scheme 2 eqn 9}, one obtains
\begin{align}
	&\alpha \sum_{i=1}^N f_i(\xb_i^{k+1}) + \alpha h(\xb_0^{k+1})-\alpha \sum_{i=1}^N f_i(\xb_i^\star)- \alpha h(\xb_0^\star)+\sum_{i=1}^N \widetilde \lambdab_i^T(\xb_i^{k+1}-\xb_0^{k+1})+\sum_{i=1}^N \frac{\alpha\sigma^2}{2}\|\xb_i^{k+1}-\xb_i^\star\|^2 \notag \\
	&~~  +\frac{1}{\beta}\sum_{i=1}^N(\widetilde \lambdab_i^{k+1}-\widetilde \lambdab_i)^T(\widetilde \lambdab_i^{k+1}-\widetilde \lambdab_i^k)
	+N\beta(\xb_0^{k+1}-\xb_0^{k})^T(\xb^{k+1}_0-\xb_0^\star)
	\notag
\end{align}
\begin{align}
	&~~+\sum_{i\in\Ac_k^c}(\widetilde \lambdab_i^{\widetilde k_i+1}-\widetilde \lambdab_i^{k+1}+\widetilde \lambdab^{\widehat k_i+1}_i-\widetilde \lambdab^{\widetilde k_i})^T(\xb_i^{k+1}-\xb_i^\star)+\sum_{i\in\Ac_k}(\widetilde \lambdab^{\bar k_i+1}_i-\widetilde \lambdab^{k}_i)^T(\xb_i^{k+1}-\xb_i^\star)\notag \\
	&~~+\sum_{i\in\Ac_k^c}\beta(\xb_0^{\widetilde k_i+1}-\xb_0^{\widehat k_i+1}-\xb_0^{k+1}+\xb_0^{k})^T(\xb_i^{k+1}-\xb_i^\star)+\sum_{i\in\Ac_k}\beta(\xb_0^{k}-\xb_0^{\bar k_i+1})^T(\xb_i^{k+1}-\xb_i^\star)\leq 0,
	\label{thm: conv of scheme 2 eqn 13}
\end{align}
where the seventh term in the LHS is obtained by applying \eqref{eqn: async cadmm s2 lambda equi2}.

We sum \eqref{thm: conv of scheme 2 eqn 13} for $k=0,\ldots,K-1$ and take the average, which yields
\begin{align}
	&\frac{\alpha}{K} \sum_{k=0}^{K-1}\bigg[\sum_{i=1}^N f_i(\xb_i^{k+1})+ h(\xb_0^{k+1})\bigg] -\alpha \bigg[\sum_{i=1}^N f_i(\xb_i^\star)+h(\xb_0^\star)\bigg]+
	\frac{1}{K} \sum_{k=0}^{K-1}\sum_{i=1}^N \widetilde \lambdab_i^T(\xb_i^{k+1}-\xb_0^{k+1}) \notag \\
	&~~~  +\frac{1}{\beta K} \underbrace{\sum_{k=0}^{K-1}\sum_{i=1}^N(\widetilde \lambdab_i^{k+1}-\widetilde \lambdab_i)^T(\widetilde \lambdab_i^{k+1}-\widetilde \lambdab_i^k)}_{\rm (a)}
	+\frac{N\beta}{K} \underbrace{\sum_{k=0}^{K-1}(\xb_0^{k+1}-\xb_0^{k})^T(\xb^{k+1}_0-\xb_0^\star)}_{\rm (b)}
	\notag
\end{align}
\begin{align}
	&\leq -\frac{1}{K} \sum_{k=0}^{K-1}\sum_{i=1}^N \frac{\alpha\sigma^2}{2}\|\xb_i^{k+1}-\xb_i^\star\|^2 \notag\\
	&~~~+\frac{1}{K} \underbrace{\sum_{k=0}^{K-1}\bigg(-\sum_{i\in\Ac_k^c}(\widetilde \lambdab_i^{\widetilde k_i+1}-\widetilde \lambdab_i^{k+1}+\widetilde \lambdab^{\widehat k_i+1}_i-\widetilde \lambdab^{\widetilde k_i})^T(\xb_i^{k+1}-\xb_i^\star)-\sum_{i\in\Ac_k}(\widetilde \lambdab^{\bar k_i+1}_i-\widetilde \lambdab^{k}_i)^T(\xb_i^{k+1}-\xb_i^\star)\bigg)}_{\rm (c)}\notag \\
	&~~~+\frac{1}{K} \underbrace{\sum_{k=0}^{K-1}\bigg(-\sum_{i\in\Ac_k^c}\beta(\xb_0^{\widetilde k_i+1}-\xb_0^{\widehat k_i+1}-\xb_0^{k+1}+\xb_0^{k})^T(\xb_i^{k+1}-\xb_i^\star) - \sum_{i\in\Ac_k}\beta(\xb_0^{k}-\xb_0^{\bar k_i+1})^T(\xb_i^{k+1}-\xb_i^\star)\bigg)}_{\rm (d)}.
	\label{thm: conv of scheme 2 eqn 14}
\end{align}
It is easy to see that term (a)
\begin{align}
	{\rm (a)}&=\frac{1}{2}\sum_{k=0}^{K-1}\bigg(\|\widetilde\lambdab_i^{k+1}-\widetilde \lambdab_i\|^2-\|\widetilde\lambdab_i^{k}-\widetilde \lambdab_i\|^2 + \|\widetilde\lambdab_i^{k+1}-\widetilde \lambdab_i^k\|^2\bigg) \notag \\
	&=\frac{1}{2}\|\widetilde\lambdab_i^{K}-\widetilde \lambdab_i\|^2-\frac{1}{2}\|\widetilde\lambdab_i^{0}-\widetilde \lambdab_i\|^2 + \frac{1}{2}\sum_{k=0}^{K-1}\|\widetilde\lambdab_i^{k+1}-\widetilde \lambdab_i^k\|^2,\label{thm: conv of scheme 2 eqn 15}
\end{align} and similarly, term (b)
\begin{align}
	{\rm (b)}&=\frac{1}{2}\sum_{k=0}^{K-1}\bigg(\|\xb^{k+1}_0-\xb_0^\star\|^2-\|\xb^{k}_0-\xb_0^\star\|^2 + \|\xb^{k+1}_0-\xb_0^{k}\|^2\bigg) \notag \\
	&=\frac{1}{2}\|\xb^{K}_0-\xb_0^\star\|^2-\frac{1}{2}\|\xb^{0}_0-\xb_0^\star\|^2 + \frac{1}{2}\sum_{k=0}^{K-1}\|\xb^{k+1}_0-\xb_0^k\|^2.
	\label{thm: conv of scheme 2 eqn 16}
\end{align}
Notice that one can bound the term $\sum_{k=0}^{K-1}\sum_{i\in\Ac_k}(\widetilde \lambdab^{\bar k_i+1}_i-\widetilde \lambdab^{k}_i)^T(\xb_i^{k+1}-\xb_i^\star)$ in (c) as follows
\begin{align}
	&\sum_{k=0}^{K-1}\sum_{i\in\Ac_k}(\widetilde \lambdab^{\bar k_i+1}_i-\widetilde \lambdab^{k}_i)^T(\xb_i^{k+1}-\xb_i^\star)
	=\sum_{k=0}^{K-1}\sum_{i\in\Ac_k}\sum_{\ell=\bar k_i+1}^{k-1}(\widetilde \lambdab^{\ell}_i-\widetilde \lambdab^{\ell+1}_i)^T(\xb_i^{k+1}-\xb_i^\star) \notag \\
	&\leq \sum_{k=0}^{K-1}\sum_{i\in\Ac_k}\sum_{\ell=k-\tau+1}^{k-1}\|\widetilde \lambdab^{\ell}_i-\widetilde \lambdab^{\ell+1}_i\|\cdot\|\xb_i^{k+1}-\xb_i^\star\| \notag\\
	&\leq \sum_{i=1}^N\sum_{k=0}^{K-1}\sum_{\ell=k-\tau+1}^{k-1}\bigg(\frac{1}{2\beta^2}\|\widetilde \lambdab^{\ell}_i-\widetilde \lambdab^{\ell+1}_i\|^2+\frac{\beta^2}{2}\|\xb_i^{k+1}-\xb_i^\star\|^2\bigg) \label{thm: conv of scheme 2 eqn 17}\\
	&\leq  \sum_{i=1}^N\sum_{k=0}^{K-1}
	\bigg(\frac{\tau-1}{2\beta^2}\|\widetilde \lambdab^{k+1}_i-\widetilde \lambdab^{k}_i\|^2+\frac{(\tau-1)\beta^2}{2}\|\xb_i^{k+1}-\xb_i^\star\|^2\bigg),\label{thm: conv of scheme 2 eqn 18}
\end{align}
where the second inequality is obtained by applying the Young's inequality:
\begin{align}
	\ab^T\bb \leq \frac{1}{2\delta}\|\ab\|^2 + \frac{\delta}{2}\|\bb\|^2
\end{align}
for any $\ab$, $\bb$ and $\delta>0$; the last inequality is caused by the fact that the term $\|\widetilde \lambdab^{k+1}_i-\widetilde \lambdab^{k}_i\|^2$ for each $k$ does not appear more than $\tau-1$ times in the RHS of \eqref{thm: conv of scheme 2 eqn 17}.
By applying a similar idea to the first term of (c) and by \eqref{thm: conv of scheme 2 eqn 18}, one eventually can bound
(c) as follows
\begin{align}
	{\rm (c)}\leq  \frac{3(\tau-1)}{2\beta^2}\sum_{i=1}^N\sum_{k=0}^{K-1}
	\|\widetilde \lambdab^{k+1}_i-\widetilde \lambdab^{k}_i\|^2+\frac{3(\tau-1)\beta^2}{2}\sum_{i=1}^N\sum_{k=0}^{K-1}\|\xb_i^{k+1}-\xb_i^\star\|^2.\label{thm: conv of scheme 2 eqn 19}
\end{align}
Similarly, the term $\sum_{k=0}^{K-1}\sum_{i\in\Ac_k}\beta(\xb_0^{k}-\xb_0^{\bar k_i+1})^T(\xb_i^{k+1}-\xb_i^\star)$ in (d) can be upper bounded as follows
\begin{align}
	&\sum_{k=0}^{K-1}\sum_{i\in\Ac_k}\beta(\xb_0^{k}-\xb_0^{\bar k_i+1})^T(\xb_i^{k+1}-\xb_i^\star)
	\leq \sum_{k=0}^{K-1}\sum_{i\in\Ac_k}\sum_{\ell=k-\tau+1}^{k-1}\beta\|\xb_0^{k}-\xb_0^{\bar k_i+1}\| \cdot\|\xb_i^{k+1}-\xb_i^\star\| \notag\\
	&\leq \sum_{i=1}^N\sum_{k=0}^{K-1}\sum_{\ell=k-\tau+1}^{k-1}\bigg(\frac{1}{2}\|\xb_0^{k+1}-\xb_0^{k}\|^2
	+\frac{\beta^2}{2}\|\xb_i^{k+1}-\xb_i^\star\|^2\bigg) \label{thm: conv of scheme 2 eqn 19.5}\\
	&\leq  \sum_{i=1}^N\sum_{k=0}^{K-1}
	\bigg(\frac{\tau-1}{2}\|\xb_0^{k+1}-\xb_0^{k}\|^2+\frac{(\tau-1)\beta^2}{2}\|\xb_i^{k+1}-\xb_i^\star\|^2\bigg).
	\label{thm: conv of scheme 2 eqn 20}
\end{align}
By applying a similar idea to the first term of (d) and by \eqref{thm: conv of scheme 2 eqn 20}, one can bound
(d) as follows
\begin{align}
	{\rm (d)}\leq  \sum_{i=1}^N\sum_{k=0}^{K-1}
	\bigg(\tau\|\xb_0^{k+1}-\xb_0^{k}\|^2+\tau\beta^2\|\xb_i^{k+1}-\xb_i^\star\|^2\bigg).\label{thm: conv of scheme 2 eqn 21}
\end{align}
After substituting \eqref{thm: conv of scheme 2 eqn 15}, \eqref{thm: conv of scheme 2 eqn 16}, \eqref{thm: conv of scheme 2 eqn 19} and \eqref{thm: conv of scheme 2 eqn 21} into \eqref{thm: conv of scheme 2 eqn 14}, we obtain that
\begin{align}
	&\alpha \bigg[\sum_{i=1}^N f_i(\bar \xb_i^{K})+h(\bar \xb_0^K)\bigg]-\alpha \bigg[\sum_{i=1}^N f_i(\xb_i^\star)
	+h(\xb_0^\star)\bigg]+
	\sum_{i=1}^N \widetilde \lambdab_i^T(\bar \xb_i^{K}-\bar \xb_0^{K}) \notag \\
	&\leq\frac{\alpha}{K} \sum_{k=0}^{K-1}\bigg[\sum_{i=1}^N f_i(\xb_i^{k+1})+h(\xb_0^{k+1})\bigg]-\alpha \bigg[\sum_{i=1}^N f_i(\xb_i^\star)+h(\xb_0^\star)\bigg]+
	\frac{1}{K} \sum_{k=0}^{K-1}\sum_{i=1}^N \widetilde \lambdab_i^T(\xb_i^{k+1}-\xb_0^{k+1})
	\notag\\
	&\leq
	\frac{1}{2\beta K}\sum_{i=1}^N\|\widetilde\lambdab_i^{0}-\widetilde \lambdab_i\|^2-\frac{1}{2\beta K}\sum_{i=1}^N\|\widetilde\lambdab_i^{K}-\widetilde \lambdab_i\|^2+\frac{N\beta}{2K}\|\xb^{0}_0-\xb_0^\star\|^2 -\frac{N\beta}{2K}\|\xb^{K}_0-\xb_0^\star\|^2
	\notag \\
	&+ \bigg(\frac{3(\tau-1)}{2K\beta^2} - \frac{1}{2\beta K}\bigg)\sum_{i=1}^N\sum_{k=0}^{K-1}\|\widetilde\lambdab_i^{k+1}-\widetilde \lambdab_i^k\|^2 +  \bigg(\frac{N\tau}{K}-\frac{N\beta}{2K}\bigg)\sum_{k=0}^{K-1}\|\xb^{k+1}_0-\xb_0^k\|^2 \notag 
\end{align}
\begin{align}
	&~~~~+\frac{1}{K} \sum_{k=0}^{K-1}\sum_{i=1}^N \bigg( \frac{3(\tau-1)\beta^2+2\tau\beta^2-\alpha\sigma^2}{2}\bigg)\|\xb_i^{k+1}-\xb_i^\star\|^2
	\label{thm: conv of scheme 2 eqn 22}
\end{align}
where the first inequality is by the convexity of $f_i$'s and $h$.

According to \eqref{thm: conv of scheme 2 eqn 22}, by choosing
\begin{align}\label{thm: conv of scheme 2 eqn 22.5}
	\beta \geq \max\{2\tau,3(\tau-1)\},~~\alpha\geq \frac{(5\tau-3)\beta^2}{\sigma^2},
\end{align} and recalling that $\lambdab_i=\widetilde \lambdab_i/\alpha$ and $\rho=\beta/\alpha$,
one can obtain
\begin{align}
	&\bigg[\sum_{i=1}^N f_i(\bar \xb_i^{K})+h(\bar \xb_0^K)\bigg]-\bigg[\sum_{i=1}^N f_i(\xb_i^\star)
	+h(\xb_0^\star)\bigg]+
	\sum_{i=1}^N \lambdab_i^T(\bar \xb_i^{K}-\bar \xb_0^{K}) \notag \\
	&\leq \frac{1}{2\rho K}\sum_{i=1}^N\|\lambdab_i^{0}-\lambdab_i\|^2+\frac{N\rho}{2K}\|\xb^{0}_0-\xb_0^\star\|^2.
	\label{thm: conv of scheme 2 eqn 23}
\end{align}
Note that \eqref{thm: conv of scheme 2 eqn 22.5} is equivalent to
\begin{align}
	\rho =\beta/\alpha \leq \frac{\sigma^2}{(5\tau-3)\beta}\leq
	\frac{\sigma^2}{(5\tau-3)\max\{2\tau,3(\tau-1)\}}.
\end{align}

Now, let $\lambdab_i=\lambdab_i^\star+\frac{\bar \xb_i^{K}-\bar \xb_0^{K}}{\|\bar \xb_i^{K}-\bar \xb_0^{K}\|}~\forall i\in \Vc$ in \eqref{thm: conv of scheme 2 eqn 23}, and note that, by the duality theory \cite{BK:BoydV04},
$$
\bigg[\sum_{i=1}^N f_i(\bar \xb_i^{K})+h(\bar \xb_0^K)\bigg]-\bigg[\sum_{i=1}^N f_i(\xb_i^\star)
+h(\xb_0^\star)\bigg]+
\sum_{i=1}^N (\lambdab_i^\star)^T(\bar \xb_i^{K}-\bar \xb_0^{K})\geq 0.
$$
Thus, we obtain that
\begin{align}
	&\sum_{i=1}^N \|\bar \xb_i^{K}-\bar \xb_0^{K}\| \leq \frac{1}{K}\bigg[\frac{1}{2\rho} \max_{\|\ab\|\leq 1}\bigg\{\sum_{i=1}^N\|\lambdab_i^{0}-\lambdab_i^\star + \ab\|^2\bigg\}+\frac{N\rho}{2}\|\xb^{0}_0-\xb_0^\star\|^2\bigg]\triangleq \frac{C_1}{K}.
	\label{thm: conv of scheme 2 eqn 23}
\end{align}
On the other hand, let $\lambdab_i=\lambdab_i^\star$ in \eqref{thm: conv of scheme 2 eqn 23}, and note that,
\begin{align}
	&\bigg[\sum_{i=1}^N f_i(\bar \xb_i^{K})+h(\bar \xb_0^K)\bigg]-\bigg[\sum_{i=1}^N f_i(\xb_i^\star)
	+h(\xb_0^\star)\bigg]+
	\sum_{i=1}^N (\lambdab_i^\star)^T(\bar \xb_i^{K}-\bar \xb_0^{K}) \notag\\
	&\geq \bigg|\bigg[\sum_{i=1}^N f_i(\bar \xb_i^{K})+h(\bar \xb_0^K)\bigg]-\bigg[\sum_{i=1}^N f_i(\xb_i^\star)
	+h(\xb_0^\star)\bigg]\bigg|-
	\delta_\lambda\sum_{i=1}^N \|\bar \xb_i^{K}-\bar \xb_0^{K}\|
\end{align} where $\delta_\lambda\triangleq \max\{\|\lambdab_1^\star\|,\ldots,\|\lambdab_N^\star\|\}$. Thus, we obtain that
\begin{align}
	&\bigg|\bigg[\sum_{i=1}^N f_i(\bar \xb_i^{K})+h(\bar \xb_0^K)\bigg]-\bigg[\sum_{i=1}^N f_i(\xb_i^\star)
	+h(\xb_0^\star)\bigg]\bigg|\notag \\
	&\leq \frac{\delta_\lambda C_1}{K}+\frac{1}{2\rho K}\sum_{i=1}^N\|\lambdab_i^{0}-\lambdab_i^\star\|^2+\frac{N\rho}{2K}\|\xb^{0}_0-\xb_0^\star\|^2=\frac{\delta_\lambda C_1 +C_2}{K}.
	\label{thm: conv of scheme 2 eqn 24}
\end{align}
Finally, combining \eqref{thm: conv of scheme 2 eqn 23} and \eqref{thm: conv of scheme 2 eqn 24} gives rise to (52).
\hfill $\blacksquare$

		\vspace{-0.0cm}
		\footnotesize
		\bibliography{distributed_opt,smart_grid,ref}
	\end{document}